\documentclass[%
reprint,
superscriptaddress,
amsmath,amssymb,
aps,longbibliography,
pra,
]{revtex4-2}

\usepackage{graphicx}%
\usepackage{verbatim}
\usepackage[version=4]{mhchem}
\usepackage{xcolor}
\usepackage{physics}
\usepackage[normalem]{ulem}
\usepackage{pdfpages}
\usepackage{pgffor}
\usepackage{hyperref}
\usepackage{tabularx}
\usepackage{soul}
\usepackage{booktabs}

\makeatletter
\AtBeginDocument{\let\LS@rot\@undefined}
\makeatother

\usepackage[normalem]{ulem}
\definecolor{tangerine}{rgb}{0.944,0.522,0}
\definecolor{verde}{rgb}{0.,0.6,0}
\definecolor{rosso}{rgb}{0.9,0.0,0.2}
\definecolor{orange}{rgb}{1.0,0.5,0.0}

\newif\ifhighlight

\newcommand{\highlight}{\highlighttrue}
\highlight
\newcommand{\editor}[2]{%
  \expandafter\newcommand\csname #1note\endcsname[1]{%
    \textcolor{#2}{(\textbf{#1note:} \textsc{##1})}}%
  \expandafter\newcommand\csname #1\endcsname[1]{%
    \ifhighlight\textcolor{#2}{##1} \else ##1\fi}%
  \expandafter\newcommand\csname #1cancel\endcsname[1]{%
    \ifhighlight\textcolor{#2}{\sout{##1}}\fi}%
  \expandafter\newcommand\csname #1change\endcsname[2]{%
    \ifhighlight\textcolor{#2}{\sout{##1} ##2}\else ##2\fi}%
  \newenvironment{#1text}{\ifhighlight\color{#2}\fi}{\color{black}}
}

\editor{MC}{red}
\editor{resub}{blue}
\editor{DT}{verde}
\editor{PP}{purple}
\editor{MK}{verde}
\editor{AM}{orange}

\begin{document}

\title{PET-MAD, a lightweight universal interatomic potential for advanced materials modeling}

\newcommand{\FB}[1] {{\color{teal} #1}}
\newcommand{\FBcancel}[1] {{\color{teal} \st{#1}}}

\author{Arslan Mazitov}
\affiliation{Laboratory of Computational Science and Modeling, Institut des Mat\'eriaux, \'Ecole Polytechnique F\'ed\'erale de Lausanne, 1015 Lausanne, Switzerland}

\author{Filippo Bigi}
\affiliation{Laboratory of Computational Science and Modeling, Institut des Mat\'eriaux, \'Ecole Polytechnique F\'ed\'erale de Lausanne, 1015 Lausanne, Switzerland}

\author{Matthias Kellner}
\affiliation{Laboratory of Computational Science and Modeling, Institut des Mat\'eriaux, \'Ecole Polytechnique F\'ed\'erale de Lausanne, 1015 Lausanne, Switzerland}

\author{Paolo Pegolo}
\affiliation{Laboratory of Computational Science and Modeling, Institut des Mat\'eriaux, \'Ecole Polytechnique F\'ed\'erale de Lausanne, 1015 Lausanne, Switzerland}

\author{Davide Tisi}
\affiliation{Laboratory of Computational Science and Modeling, Institut des Mat\'eriaux, \'Ecole Polytechnique F\'ed\'erale de Lausanne, 1015 Lausanne, Switzerland}

\author{Guillaume Fraux}
\affiliation{Laboratory of Computational Science and Modeling, Institut des Mat\'eriaux, \'Ecole Polytechnique F\'ed\'erale de Lausanne, 1015 Lausanne, Switzerland}

\author{Sergey Pozdnyakov}
\affiliation{Laboratory of Computational Science and Modeling, Institut des Mat\'eriaux, \'Ecole Polytechnique F\'ed\'erale de Lausanne, 1015 Lausanne, Switzerland}

\author{Philip Loche}
\affiliation{Laboratory of Computational Science and Modeling, Institut des Mat\'eriaux, \'Ecole Polytechnique F\'ed\'erale de Lausanne, 1015 Lausanne, Switzerland}

\author{Michele Ceriotti}
\email{michele.ceriotti@epfl.ch}
\affiliation{Laboratory of Computational Science and Modeling, Institut des Mat\'eriaux, \'Ecole Polytechnique F\'ed\'erale de Lausanne, 1015 Lausanne, Switzerland}

\date{\today}%

\begin{abstract}
Machine-learning interatomic potentials (MLIPs) have greatly extended the reach of atomic-scale simulations, offering the accuracy of first-principles calculations at a fraction of the cost. 
Leveraging large quantum mechanical databases and expressive architectures, recent ``universal'' models deliver qualitative accuracy across the periodic table but are often biased toward low-energy configurations.
We introduce PET-MAD, a generally applicable MLIP trained on a dataset combining stable inorganic and organic solids, systematically modified to enhance atomic diversity. 
Using a moderate but highly-consistent level of electronic-structure theory, we assess PET-MAD’s accuracy on established benchmarks and advanced simulations of six materials. 
Despite the small training set and lightweight architecture, PET-MAD is competitive with state-of-the-art MLIPs for inorganic solids, while also being reliable for molecules, organic materials, and surfaces. It is stable and fast, enabling the near-quantitative study of thermal and quantum mechanical fluctuations, functional properties, and phase transitions out of the box. It can be efficiently fine-tuned to deliver full quantum mechanical accuracy with a minimal number of targeted calculations.
\end{abstract}

\maketitle

\newlength{\halfpage}
\setlength{\halfpage}{\columnwidth}

\section{Introduction}\label{sec:introduction}
Efficient methods for solving the electronic structure problem for molecules and materials have made it possible to predict their structure, stability and properties from first principles. 
However, the associated computational cost and poor scaling with system size severely limit the complexity, time, and length scale accessible to simulations.
Machine-learning (ML) models that aim to replace first-principles methods to evaluate energy and forces --- so-called machine-learning interatomic potentials (MLIPs) --- address this limitation by training on a set of reference quantum mechanical calculations, and then making accurate yet inexpensive ML predictions, for both materials~\cite{behl-parr07prl,bart+10prl,deri+19am} and (bio)molecules \cite{rupp+12prl,smit+17cs,wang+19acscs,seut+25cs}.

For almost two decades, this scheme has been used successfully to study a large number of chemical systems for which first-principles calculations are too expensive, and empirical forcefields are either not available or not sufficiently accurate~\cite{zuo+20jpcl, unke+21cr,nong+21nc}.
Nonetheless, the aforementioned approach typically implies parameterizing a specific MLIP for every new system, which requires a considerable number of \textit{ab initio} calculations, as well as substantial human effort and expertise in fitting the potential. 
In contrast to this family of single-purpose potentials, recent years have seen the development of several general-purpose, or ``universal'' models \cite{Batatia2023MACE-MP-0, Yang2024Mattersim, Neumann2024},  which aim to be applicable, either out of the box or after minimal fine-tuning, to a large range of distinct chemical systems.
These models provide a qualitative description of the atomic-scale interactions across the periodic table, although their level of accuracy for a specific application depends on how closely the problem of interest aligns with the considerations underlying the construction of the reference dataset. For instance, many existing datasets are built to represent a collection of stable materials, and to support the search for new materials with optimal properties \cite{Deng2023, WBM-dataset, Wang2023, Merchant2023}. 
To complicate things, the accuracy of general-purpose potentials is often assessed using inconsistent levels of \textit{ab initio} theory, making it difficult to isolate the shortcomings of the model from the discrepancy in the electronic-structure reference.
In this work, we use the Massive Atomistic Diversity (MAD) dataset, 
which incorporates a high degree of chemical and structural diversity while employing highly converged, internally consistent reference calculations. 
This allows for a quantitative assessment of the performance of fitted models across the periodic table, since MAD's \textit{ab initio} settings are broadly applicable to most chemical systems, although the comparatively simple exchange-correlation functional we use may not always be equally accurate. 
Energies and forces are fitted using a \textit{Point Edge Transformer} (PET) graph neural network~\cite{pozd-ceri23nips}, resulting in a general-purpose MLIP, which we name PET-MAD. 
We evaluate the accuracy of the resulting  model on a wide range of public datasets, ensuring consistency in the  electronic-structure settings, and showing that PET-MAD achieves competitive performance on several external benchmarks despite being trained on 2-3 orders of magnitude fewer structures. 
We then critically assess the performance of PET-MAD in highly non-trivial atomistic simulations, including accelerated statistical sampling, quantum nuclear fluctuations, and predicting functional properties. 
We focus on six examples, motivated by their scientific interest, diversity in material classes and physical effects being probed, and the fact that each has previously been studied using \emph{ad hoc} MLIPs --- namely, lithium thiophosphate, gallium arsenide, a CoCrFeMnNi high-entropy alloy, liquid water, succinic acid, and barium titanate. 
For each example, we quantitatively assess the accuracy of PET-MAD by comparing it against both a bespoke model trained with compatible energetics, that provides an extremely close match to the electronic-structure reference, as well as a finetuned model that enables PET-MAD to achieve the same level of accuracy with a small number of additional training structures. PET-MAD demonstrates that accurate, fast, and robust universal models can be trained using a tiny fraction of the structures of last-generation datasets, and provides a feature-rich framework for advanced atomistic simulations, which includes direct-force acceleration and inexpensive end-to-end uncertainty quantification.

\section{Results}\label{sec:results}
PET-MAD is a generally-applicable machine-learning interatomic potential based on PET (an unconstrained, transformer-based graph neural network architecture) and a custom training set built on the principles of Massive Atomistic Diversity (MAD) and internal consistency of the reference energetics.
We thoroughly test the performance of PET-MAD against several benchmark datasets and compare it with four widely used universal potentials. 
Six case studies complete our analysis, showcasing and quantitatively assessing PET-MAD for diverse classes of materials and advanced atomistic simulations, comparing a system-specific PET versions trained on an \emph{ad hoc} datasets, the generally-applicable PET-MAD, and the versions finetuned to each system. 
We emphasize that in all of these examples we focus on the consistency of the model with the electronic-structure reference, irrespective of the actual accuracy of the latter in reproducing experimental quantities.

\subsection{MAD dataset}
\label{sec:dataset_details}

Most of the existing efforts to generate datasets to train universal models focus on either inorganic crystals~\cite{Deng2023, Schmidt2023,Wang2023} or molecular compounds~\cite{east+23sd}, and aim to include as many structures as possible.
PET-MAD is trained on the Massive Atomistic Diversity (MAD) dataset, 
which is based on a different philosophy.
First, it aims to push the limits of ``universality'' by including both organic and inorganic atomistic systems of all possible dimensionalities. This approach has been already shown to deliver good transferability upon including up to 45 elements in the training data \cite{takamoto2022}.
Second, our resulting model is meant to work in complex atomistic simulation protocols describing a wide range of thermodynamic conditions, which requires covering a large configuration space and being computationally affordable. To this end, the MAD dataset includes randomized and highly-distorted structures in the training set, applying to solids and surfaces ideas similar to those that drove the development of ``mindless'' molecular benchmark sets \cite{Korth2009}.
Third, the reference electronic-structure calculations are designed to be robust and consistent, to ensure that the structural and chemical motifs are treated in the same way, regardless of the type of structure they are part of, which is important for a coherent structure-energy mapping. 
This choice neglects the description of effects such as spin polarization, electron correlation, and dispersion, which are important for certain classes of materials but cannot be applied consistently across the MAD dataset.
Details of the electronic-structure calculations are given in Section \ref{sec:dft_details}.
Last but not least, we restrict the dataset to fewer than 100,000 structures in order to reduce the time and cost of training, making the entire training procedure accessible to a wider community. As we shall see, the strategy we use allows us to preserve the representative power of the dataset. 

The MAD dataset contains 95595 structures of 85 elements in total (with atomic numbers ranging from 1 to 86, except Astatine) and consists of the following 8 subsets:

\begin{description}
\item[\textbf{MC3D}] Bulk crystals from the Materials Cloud 3D crystals database \cite{Huber2022MC3D} (33596 structures)
\item[\textbf{MC3D-rattled}] Rattled analogs of the original MC3D crystals, with Gaussian noise added to all atomic positions (30044 structures)
\item[\textbf{MC3D-random}] Artificial structures obtained by replacing the atomic species of a few MC3D structures with a random sampling from the list of all 85 elements (2800 structures) 
\item[\textbf{MC3D-surface}] Surface slabs generated from the MC3D structures by cleaving the crystal along a random crystallographic plane with low Miller index (5589 structures)
\item[\textbf{MC3D-cluster}] Nanoclusters, generated by cutting a random atomic environment of 2 to 8 atoms from a random crystal sampled from the MC3D and MC3D-rattled subsets (9071 structures)
\item[\textbf{MC2D}] Two-dimensional crystals from the Materials Cloud 2D crystals database  \cite{Campi2023MC2D} (2676 structures)
\item[\textbf{SHIFTML-molcrys}] A curated subset of the SHIFTML molecular crystals structures \cite{Cordova2022}, that are in turn sampled from the Cambridge Structural Database \cite{groom_cambridge_2016} including both relaxed and thermally-distorted configurations   (8578 structures)
\item[\textbf{SHIFTML-molfrags}] Neutral molecular fragments extracted from the SHIFTML dataset \cite{Cersonsky2023} (3241 structures)
\end{description}

The number of structures in each subset represents the total number of structures that were successfully converged in DFT calculations, after filtering out a few outlier configurations with forces above a large threshold (100 eV/Å\ for MC3D-rattled and MC3D-random, and 50 eV/Å\ for the other subsets). Specific details on the generation of MAD subsets are given in Section \ref{sec:mad_details}. Details of the first principles calculations are provided below in Section \ref{sec:dft_details}. 

\subsection{Model architecture and training}\label{sec:model_architecture}

The PET architecture \cite{pozd-ceri23nips} is a rotationally unconstrained and transformer-based graph neural network (GNN) which has a high descriptive power (as every transformer layer can be scaled to be a universal approximator) \cite{pozd-ceri23nips} and low inference cost (as we shall demonstrate in Fig.~\ref{fig:speed}). 
For PET-MAD, we chose the architecture based on an extensive hyperparameter search, resulting in a model with 
approximately 3.3 million parameters in total. 
Before training, each subset of MAD was randomly shuffled and further split into training, validation, and test subsets in fractions of 80\%, 10\%, and 10\%, respectively. All subsets were then merged to obtain the final training, validation, and test sets. More details on the architecture and training are available in 
Section \ref{sec:pet_details}, Section \ref{sec:pet-mad_details}, and the Supplementary Information. In order to improve the performance of PET-MAD on specific chemical systems, we used a fine-tuning strategy based on the low-rank adaptation technique (LoRA)~\cite{LoRA}. Details on fine-tuning are further discussed in Section \ref{sec:fine-tuning-details}.

The model also includes uncertainty-quantification capabilities, based on the last-layer prediction rigidity (LLPR) method~\cite{bigi+24mlst}.
Error estimation is especially important for a model aiming to be generally-applicable, and that (despite the high diversity of the MAD training set) may often be used  for out-of-domain predictions. As shown in Supplementary Figure \ref{fig:pet-mad-uncertainties}, the calibrated LLPR uncertainties correlate well with the empirical test-set errors, while adding negligible overhead on top of energy predictions. 
Given that PET-MAD is intended for use in advanced simulations that estimate observable quantities through sophisticated sampling protocols, it is essential to be able to propagate the energy errors onto the final quantity of interest. 
To this end, we generate a shallow ensemble based on the LLPR covariance~\cite{bigi+24mlst}.
Unlike standard ensemble-based methods \cite{Lakshminarayanan2017}, which typically introduce a computational overhead proportional to the number of ensemble members, both LLPR and the shallow ensemble method~\cite{kell-ceri24mlst} allow uncertainty quantification at negligible additional cost. Moreover, they allow one to compute errors on any derived quantities using direct ensemble propagation, as we demonstrate for free-energy calculations in Section \ref{sec:GaAs_result}, and for phonon dispersion curves in Section~\ref{sec:phonons}.
Details on our uncertainty-quantification approach are summarized in Section \ref{sec:uq} and explained thoroughly in the Supplementary Information.
Even though PET-MAD is designed to evaluate forces using backpropagation, we also train a separate head that predicts forces directly as a function of the atomic coordinates. 
Avoiding backpropagation makes the model two to three times faster (Figure~\ref{fig:speed}), but should be used with care, as it violates energy conservation with pathological consequences on sampling accuracy~\cite{Bigi2024}. 
We address these problems in Section~\ref{sec:directforces}, where we also show how a multiple-time-step integrator allows one to retain the computational advantages of direct force predictions without introducing sampling artifacts.

\begin{figure*}
    \includegraphics[width=0.75\textwidth]{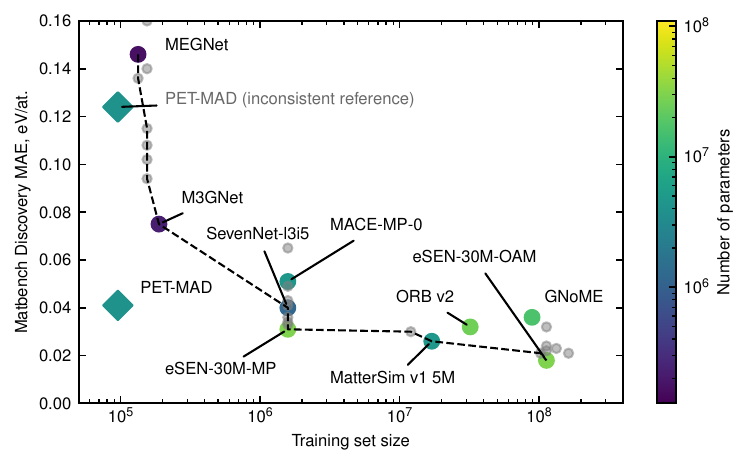}
    \caption{\label{fig:umlip_frontier}
    Pareto frontier of various universal MLIPs: x-axis shows the number of configurations in the training set, and y-axis shows mean absolute error (MAE) for the energy-above-hull property on a subset of the WBM dataset \cite{WBM-dataset} (a part of the Matbench Discovery benchmark \cite{Matbench}). Colored dots represent a set of selected models, while other models from the benchmark are shown with grey dots. PET-MAD results are shown with diamonds for both consistent and inconsistent DFT references. PET-MAD significantly pushes the frontier downward, achieving the accuracy of other advanced universal MLIPs like SevenNet and GNoME using 1-3 orders of magnitude less data.}
    
\end{figure*}

\subsection{Benchmarking}
\label{sec:benchmarking}

First, we demonstrate the efficacy of combining the MAD dataset and the PET architecture by comparing the accuracy of PET-MAD against other universal models on the popular Matbench Discovery benchmark \cite{Matbench}.
We analyze the overall gain in models' accuracies with the increase in the training set size and the number of trainable parameters. 

Since PET-MAD is trained on the MAD dataset, which has different DFT settings compared to the Matbench (which shares common settings with the training sets of other reference models), we recomputed a random subset of the benchmark containing 555 structures (excluding lanthanides and actinides) using MAD-compatible DFT settings. 
We evaluate the accuracies of all models on this subset, picking for each model the DFT reference compatible with its training set.
Figure \ref{fig:umlip_frontier} demonstrates the Pareto frontier of the models' data efficiency: Pareto-optimal models efficiently use additional training data to increase their accuracy. PET-MAD significantly improves on existing models, achieving the accuracy of other advanced universal MLIPs like SevenNet and GNoME using 1-3 orders of magnitude less data. 
For illustrative purposes, the figure also shows the accuracy of PET-MAD evaluated against the inconsistent Matbench energetics: the $3\times$ larger error is entirely due to the discrepancy in the reference data. A thorough analysis of DFT consistency effects is discussed in the Supplementary Section \ref{sec:matbench-discovery}

\begin{table*}
  \centering\small
  \caption{\label{tab:benchmarks}
    Comparison of PET-MAD accuracies on popular atomistic machine learning datasets against MACE-MP-0-L, MatterSim-5M, Orb-v2, and SevenNet-l3i5 models.
  For each dataset and model, mean absolute errors are reported for raw energy$|$forces predictions (in units of meV/atom$|$meV/Å). We do not show force errors for the Matbench Discovery data, since they are not available in the reference data.}
  \begin{tabular}{lccccc} \toprule
    Dataset      & PET-MAD     & MACE-MP-0-L & MatterSim-5M & Orb-v2      & SevenNet \\ \midrule
    MAD          & \textbf{17.6}$|$\textbf{65.1}& 81.6$|$181.5  & 47.3$|$133.7   & 52.9$|$96.2& 82.1$|$173.5 \\
    MPtrj        & 22.3$|$77.9& 15.1$|$50.8& 21.3$|$61.4& \textbf{5.6}$|$\textbf{21.9}& 9.8$|$25.5\\
    Matbench     & \textbf{31.3}$|$ --- & 58.5$|$---    & 38.2 $|$---     & 37.9$|$---  & 47.5$|$---\\
    Alexandria       & 49.0$|$66.8& 65.4$|$79.5& 21.2$|$39.9& \textbf{13.2}$|$\textbf{10.5}& 47.6$|$70.3\\
     OC2020      & \textbf{18.3}$|$114.5& 82.4$|$169.6& 31.5$|$119.2& 19.8$|$\textbf{99.3}& 45.7$|$162.7\\
    SPICE    & \textbf{3.7}$|$\textbf{59.5}& 10.6$|$166.8& 21.3$|$145.6& 59.0$|$140.8& 11.3$|$139.1\\
    MD22         & \textbf{1.9}$|$\textbf{65.6}& 9.4$|$182.9& 28.6$|$160.4& 174.3$|$220.7& 11.1$|$146.2\\
    \bottomrule
  \end{tabular} 
\end{table*}

The figure also shows that many of the optimal models use a very large number of parameters, which makes them accurate but  computationally demanding, and therefore not suitable for the long, complex simulation tasks that PET-MAD is designed for.
For this reason, we will focus on medium-size models for the rest of these benchmarks: MACE-MP-0 \cite{Batatia2023MACE-MP-0},  MatterSim \cite{Yang2024Mattersim}, Orb-v2 \cite{Neumann2024} and SevenNet \cite{Park2024}.  Like PET-MAD, these models have been designed and are widely used for advanced materials simulations beyond single-point calculations.
We extend the comparison on Matbench to include other popular atomistic machine learning benchmarks for bulk inorganic systems, molecular systems, and catalytic applications. Thus, we use a total of seven following benchmarks: 
\begin{description}
    \item[\textbf{MAD}] The dataset developed in this work
    \item[\textbf{MPtrj}] Relaxation trajectories of bulk inorganic crystals dataset from Ref. \citenum{Deng2023}
    \item[\textbf{Matbench}] The dataset of single-element substitutions on bulk inorganic crystals from Ref. \citenum{WBM-dataset}
    \item[\textbf{Alexandria}] Relaxation trajectories of bulk inorganic crystals as well as 2D, 1D systems from Ref.~\citenum{Schmidt2023}
    \item[\textbf{OC2020 (S2EF)}] Molecular relaxation trajectories on catalytically active surfaces from Ref.~\citenum{Chanussot2021}
    \item[\textbf{SPICE}] Drug-like molecules and peptides from Ref.~\citenum{east+23sd}
    \item[\textbf{MD22}] Molecular dynamics trajectories of peptides, DNA molecules, carbohydrates and fatty acids from Ref.~\citenum{Chmiela2023}
\end{description}
In evaluating the benchmarks, it is important to keep in mind that (1) most of the reference models are larger than PET-MAD (we use the versions  with the largest number of parameters in their families, MACE-MP-0-L and MatterSim-5M, Orb-v2,  SevenNet-l3i5 having respectively about 15.8 M, 4.6 M, 25 M, and 1.17 M parameters, as opposed to 3.3 M for PET-MAD) and (2) they are trained on much larger datasets (1.58 M, 17 M, 32.1 M, and 1.58 M structures, respectively).

The benchmarking results are presented in Table \ref{tab:benchmarks}. 
We find that PET-MAD achieves high accuracy in predicting both energies and interatomic forces, outperforming MACE-MP-0-L in most cases, and competing closely with MatterSim-5M and SevenNet-l3i5 even for the inorganic datasets they are trained on. Upon using a consistent level of DFT theory, PET-MAD outperforms all other models in predicting raw energies of the WBM crystals from the Matbench Discovery benchmark \cite{Matbench}.
Orb-v2 performs significantly better than all other models on MPtrj and Alexandria, which are part of its training set.
PET-MAD outperforms all other models on molecular datasets such as SPICE and MD22, and matches Orb-v2 also on the OC2020 S2EF -- which is a strongly extrapolative exercise as MAD does not include configurations of adsorbed molecules.

\begin{figure*}
    \includegraphics[width=\textwidth]{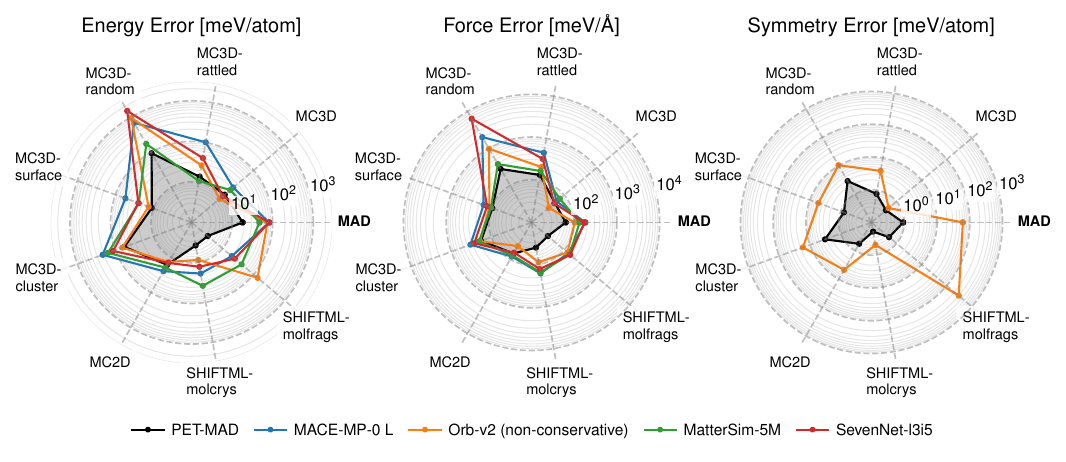}
    \caption{\label{fig:mad_benchmark}
    Comparison of accuracies of PET-MAD (black line), MACE-MP-0-L (blue line), Orb-v2 (orange line), MatterSim-5M (green line), and SevenNet-l3i5 (red line) models on various subsets of the MAD dataset. The left and center panels show the mean absolute errors in energies and force predictions. The right panel shows the rotational discrepancy in energy predictions of the unconstrained models (PET-MAD and Orb-v2). 
    Results for all models except PET-MAD were obtained using consistently recalculated reference values (i.e. computed with MPtrj dataset settings) to achieve consistency with the training sets of the corresponding reference models.
    See also Table \ref{tab:benchmarks} for benchmarks on out-of-sample datasets.
    }
\end{figure*}

The reference models perform much worse than PET-MAD on the (consistently recomputed) MAD benchmark. This is not too surprising, as MAD is designed to be more diverse than the datasets, on which these models are trained. 
Still, it is instructive to assess separately the accuracy of the various models on the different sections of MAD (Figure \ref{fig:mad_benchmark}). 
PET-MAD outperforms MACE-MP-0-L, MatterSim-5M and SevenNet-l3i5 in almost all cases. 
The accuracy is similar (and in a few cases slightly better) for the stable inorganic structures including bulk (MC3D) and layered (MC2D) compounds. MC3D-derived surfaces and clusters show slightly larger force and considerably larger energy errors.
Errors of the other universal models are up to 50 times larger for the distorted (MC3D-rattled and MC3D-randcomp) that contain especially unusual, diverse configurations. 
MatterSim-5M, which is trained on a broader, yet unpublished, set of configurations, performs substantially better than the other models.
The most notable difference occurs in the molecular systems (SHIFTML-molcrys and SHIFTML-molfrags), where PET-MAD dramatically outperforms all other models. This result is again expected, as organic systems represent a completely different region of configuration space, which is heavily undersampled in the case of inorganic crystal datasets. 
The last model from the reference list, Orb-v2, requires a separate discussion, as it provides by default direct, non-conservative force predictions. 
Orb-v2 outperforms all other models (including, marginally, PET-MAD) on relaxed inorganic systems, presumably thanks to the combination of a flexible, unconstrained architecture and the large training set (which includes both MPtrj \cite{Deng2023} and Alexandria \cite{Schmidt2023} datasets). However, it still provides worse accuracy on rattled, random composition structures, surfaces and molecular systems compared to PET-MAD. 
Both PET-MAD and Orb-v2 use architectures that do not enforce exact rotational equivariance, which is learned approximately by data augmentation. 
The resulting symmetry breaking can be monitored and controlled easily~\cite{lang+24mlst}, but it is important to assess its extent.
To do so, we estimated the symmetry error in PET-MAD and Orb-v2 energy predictions by applying a series of rotations on a Lebedev-Laikov grid of order 9 for each structure in the MAD subset and calculating the standard deviation in the predictions (Figure \ref{fig:mad_benchmark}, right panel). 
In most cases, the rotational discrepancy of the PET-MAD model is one or two orders of magnitude smaller than the corresponding prediction error, below 1 meV/atom for all subsets except MC3D-random and MC3D-cluster. Orb-v2 shows significantly higher discrepancies in energy predictions, which are at times comparable to the actual errors.

\begin{figure*}
    \includegraphics[width=\textwidth]{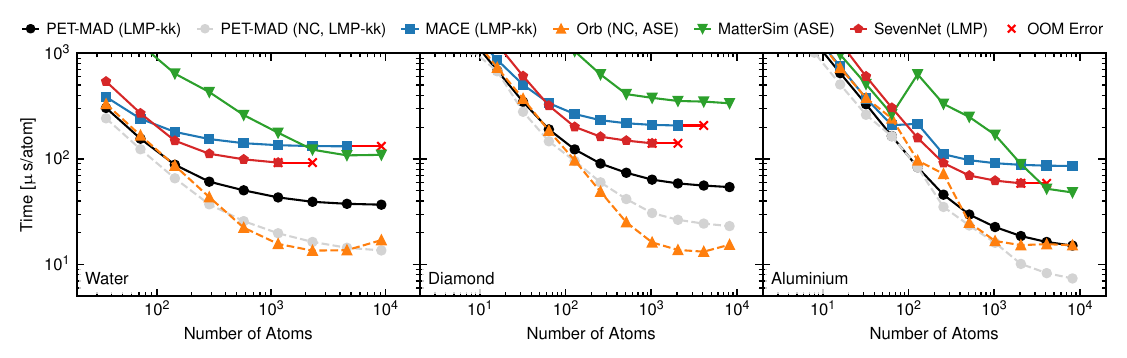}
    \caption{\label{fig:speed}
Inference time of several models evaluated over different bulk materials and system sizes. 
For each MLIP, we use its LAMMPS~\cite{plim95jcp} interface if available, preferably choosing the Kokkos-enabled \cite{KOKKOS} (kk) version, or its ASE~\cite{hjor+17jpcm} interface otherwise. 
All timings were measured on a single NVIDIA H100 GPU. Missing points for MACE-MP-0 and SevenNet caused by out-of-memory errors are marked with red crosses. The non-conservative (NC) versions of Orb-v2 and PET-MAD are shown in dashed lines. These models benefit from a theoretical speedup, but can violate conservation of energy, often resulting in ill-behaved molecular dynamics \cite{Bigi2024}. The model versions used were MatterSim-v1.0.0-1M, MACE-MP-0 (M), Orb-v2, SevenNet-0 (11Jul2024).
}
\end{figure*}

The PET-MAD model is also computationally efficient, which facilitates simulating larger time and length scales. 
We compared the throughput of molecular dynamics simulations of systems of different sizes and densities (solid aluminum, diamond, and liquid water, see Figure \ref{fig:speed}) and, when available, used lighter versions of the reference models compared to those used in the accuracy benchmarks (namely, MACE-MP-0 M, MatterSim-1M, and SevenNet-0). 
These models have fewer parameters and thus allow faster inference at the expense of accuracy, making this a more challenging test for the speed of PET-MAD. Nevertheless, we see that PET-MAD is faster and more memory efficient than all models.
When exploiting the computational advantage of direct-force prediction, PET-MAD is also competitive with the non-conservative Orb-v2 model.

The takeaway is that an unconstrained architecture and a problem-agnostic construction of the training set make PET-MAD competitive in speed and accuracy with models that have a larger number of parameters and are trained on much larger datasets (Fig.~\ref{fig:umlip_frontier}).
This has important implications to guide the design of future datasets, and provides a more sustainable and accessible platform to support model development. 

Benchmarking accuracies, however, do not solely ensure that PET-MAD can be reliably used in realistic atomistic modeling scenarios. Therefore, in the following sections, we demonstrate applicability of PET-MAD for atomistic simulations by comparing its performance against single-purpose and finetuned models for six diverse and challenging use cases.


\subsection{Ionic transport in lithium thiophosphate}

\label{sec:LPS_result}
Lithium thiophosphates (LPS) are a class of materials that have been intensely studied as electrolytes for solid-state batteries~\cite{Kwade2018}. 
The archetypal member of this class, \ce{Li3PS4}, has been the subject of several computational investigations,
including Ref.~\citenum{Gigli2024}, which we use both as a blueprint for this benchmark and the source of the system-specific dataset. 
Following the same approach as in that reference, we compute the ionic conductivity $\sigma$ (one of the key properties for the technological applications of LPS) of three known phases of \ce{Li3PS4}, $\alpha$, $\beta$ and $\gamma$~\cite{kaup_impact_2020}, using molecular dynamics and the Green-Kubo theory of linear response~\cite{Kubo}.

\begin{figure}
    \centering\includegraphics{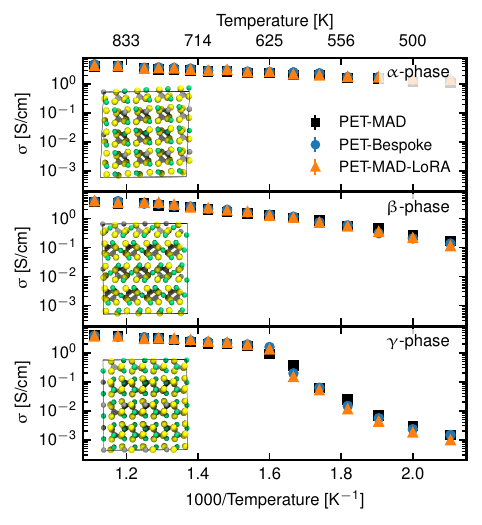}
    \caption{\label{fig:LPS-GK}
    Temperature dependence of the ionic conductivity $\sigma$ for the $\alpha$, $\beta$ and $\gamma$ phases of \ce{Li3PS4}. The figure compares the values of $\sigma$ from PET-MAD (black), a finetuned model (orange) and a bespoke PET model trained from scratch (blue).
    }
\end{figure}

We compare the results obtained using the PET-MAD potential with a single-purpose PET potential trained from scratch over the dataset from Ref.~\citenum{Gigli2024} (PET-Bespoke), and with one finetuned over the same dataset using LoRA technique (see Section \ref{sec:fine-tuning-details} for more details, and the SI for a full discussion).
For this dataset, the validation mean absolute error (MAE) of PET-MAD for energy\textbar~forces  is 4.9~meV/atom\textbar63.9~meV/Å, to be compared with 1.2~meV/atom\textbar35.6~meV/Å, for a model trained from scratch on the LPS dataset, and 1.3~meV/atom\textbar36.0~meV/Å  \ for the finetuned model. 
Figure \ref{fig:LPS-GK} shows the values of the conductivity for all the phases and over a wide range of temperatures computed with PET-MAD (black), a model finetuned using the LoRA algorithm with rank 8 (orange) and the PET-Bespoke model (blue).
For all three phases, PET-MAD is in excellent agreement with the bespoke and finetuned models, demonstrating quantitative accuracy despite the larger validation error.
For the $\gamma$-phase, PET-MAD slightly overestimates the temperature at which the transition to a high-conductivity phase occurs.
Except for this minor discrepancy, PET-MAD shows excellent agreement with the more accurate dedicated models, capturing the change of behavior in $\sigma$ following the rotation of the PS$_4$ tetrahedra~\cite{Gigli2024}.
The results agree quantitatively with those of Ref.~\citenum{Gigli2024}, despite being based on slightly different DFT parameters and relied on a different class of MLIPs. 

\begin{figure}
    \centering\includegraphics{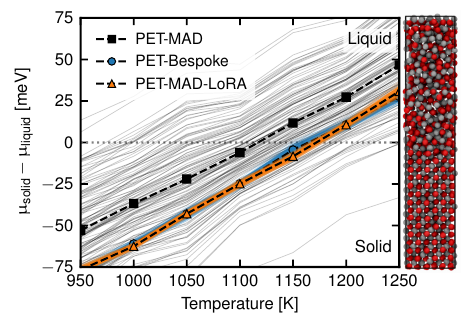}
    \caption{\label{fig:Chem_pot_curves_GaAs_UQ}
    Chemical potential differences between liquid and solid GaAs phases as a function of temperature computed with PET-MAD (black), PET-Bespoke (blue) and PET-MAD-LoRA (orange) models. The dashed solid lines indicate the mean values of the predicted $\Delta \mu$, and the light colored lines represent the reweighted predictions from the individual members of the shallow ensemble. The large spread of the PET-MAD ensemble predictions correctly reflects the discrepancy with respect to the Bespoke and LoRA-finetuned model.}
\end{figure}

\subsection{Melting point of GaAs}
\label{sec:GaAs_result}
Gallium Arsenide (GaAs) is a III-V semiconductor whose properties make it a good choice to manufacture the high-quality optical and electrical devices, as well as the high-end photovoltaics.
The growth of GaAs nanostructures often relies on the coexistence of solid GaAs and molten Ga in an As-rich atmosphere \cite{i2015semiconductor}. Corresponding computational studies typically require MLIPs that can accurately describe different phases and compositions across the Ga/As phase diagram. 
Electronic properties change wildly between different phases, posing additional challenges to the construction of empirical potentials.
Following the work of Imbalzano et al.~\cite{imba-ceri21prm}, we repeat the calculation of the melting point of stoichiometric GaAs at ambient pressure with the interface pinning method. 
This method computes the differences in chemical potentials between the liquid and solid phases, and the melting point is determined by identifying the temperature at which the chemical potential difference becomes zero. 
A combination of the Laplace ensemble predictions with the reweighting technique allows us to propagate errors on the chemical potential curves\cite{imba+21jcp}, and therefore assess the contribution of the epistemic part of the error to the melting point error caused by limitations of the ML fit.

Figure \ref{fig:Chem_pot_curves_GaAs_UQ} demonstrates these calculations, comparing the PET-MAD model with a bespoke PET model trained on the reference GaAs training set, and the LoRA-finetuned model on the same training set. The test set errors (MAE) on energies\textbar~forces are 14.4~meV/at.\textbar74.1~meV/Å\, for PET-MAD,  0.7~meV/at.\textbar29.0~meV/Å\, for PET-Bespoke and 1.3~meV/at.\textbar45.3~meV/Å\, for PET-MAD-LoRA. 
When it comes to the estimation of the melting point, the LoRA and Bespoke models are in quantitative agreement ($1169\pm 3$~K vs $1169 \pm 4$~K) whilst the PET-MAD model predicts a slightly smaller value of $1111\pm 72$~K. The predicted error is consistent with the lower accuracy of the general-purpose PET-MAD, which highlights the importance of the inexpensive uncertainty propagation afforded by the Laplace ensemble architecture implemented in our PET models. 
It should be noted that the error of the general-purpose model against the bespoke potentials is very small compared to the deviation of the computed melting point from the experimentally measured melting point (1511~K), and with the typical errors of empirical forcefields. 
This large discrepancy is consistent with that observed in Ref.~\citenum{imba-ceri21prm}, and can be attributed to shortcomings of the reference electronic-structure calculations -- which is not uncommon as the computed melting point of a given material can vary by hundreds of Kelvin depending on the choice of DFT functional\cite{jinnouchiOntheflyMachineLearning2019b}. 

\begin{figure}
    \centering\includegraphics{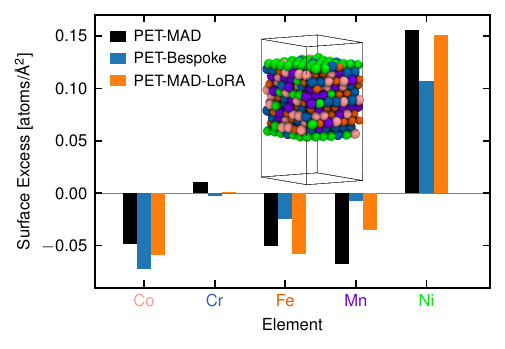}
    \caption{\label{fig:hea_surface_excess}
    Gibbs surface excess $\Gamma_a$ of the elements on a (111) surface of the CoCrFeMnNi alloy at 800 K from the REMD/MC simulation performed with the pre-trained PET-MAD model (black bars), bespoke PET model trained on a subset of the HEA25S dataset (blue bars), and the PET-MAD-LoRA model, finetuned on the same subset (orange bars). Both pre-trained and finetuned models result in almost identical segregation pattern with the surface enriched in nickel, which is similar to one from Ref. \citenum{mazi+24jpm}. The inconsistency between the bespoke and finetuned models is most likely due to the bespoke model's overfitting.}
\end{figure}

\subsection{Surface segregation in high-entropy alloys}

High-entropy alloys -- containing 5 or more metals in roughly equimolar composition -- often possess excellent mechanical properties \cite{
cant+04msea}, and perform well as heterogeneous catalysts \cite{Sun2021}. 
Investigating their properties and exploring efficiently their composition space requires versatile models that can handle a high degree of chemical diversity. We replicate some of the simulations in Ref.~\citenum{mazi+24jpm},  using PET-MAD to study the segregation of different elements at the (111) surface of the CoCrFeMnNi alloy, a prototypical HEA \cite{cant+04msea}. Modeling of the differential surface propensity of the elements in the alloy --- which is central to the applications of HEAs as heterogeneous catalysts --- requires a combination of replica-exchange molecular dynamics with Monte-Carlo atom swaps (REMD/MC) to overcome the slow diffusivity, and the existence of free-energy barriers to segregation starting from a random alloy. 
This example allows us to demonstrate the capabilities of PET-MAD in complex computational workflows, to assess the ability of the model to describe surface effects that are often overlooked in other universal models \cite{Focassio2024}, and to test finetuning over a dataset with a greater degree of chemical diversity than the other examples we consider. 
We use the Gibbs surface excess calculated at 800 K using the REMD/MC protocol  (see Section \ref{sec:hea_details} for computational details) as a measure of surface segregation in a CoCrFeMnNi surface slab. 
We compare three models: the pre-trained PET-MAD, PET trained from scratch (PET-Bespoke) and the finetuned model (PET-MAD-LoRA), where the last two were trained on a subset of the HEA25S dataset \cite{mazi+24jpm}, recomputed with MAD dataset DFT settings for consistency (details on the subset selection are given in the Supplementary Section \ref{sec:hea_details}). The validations MAEs of all three models on energy\textbar forces are 25.8 meV/atom\textbar175.1 meV/Å, 14.6 meV/atom\textbar138.3 meV/Å, and 9.4~meV/atom\textbar124.8 meV/Å, respectively.

Results on surface excess calculation are given in Figure \ref{fig:hea_surface_excess}. First, we note that the segregation pattern obtained from both pre-trained and finetuned models is almost identical and quantitatively matches the results from Ref. \citenum{mazi+24jpm}, which corresponds to surface layers enriched in nickel and depleted in other elements.
Both these models agree qualitatively with the results of the PET-Bespoke model, but there is a more pronounced quantitative difference in the surface excess values. 
In interpreting this discrepancy, one should keep in mind that, contrary to other examples, we only recomputed about 2000 structures out of the 30'000 in the HEA25S, and so a bespoke model incurs a high risk of overfitting -- which is also consistent with the learning curves reported in the SI. This case study demonstrates that pre-trained PET-MAD is capable of capturing all effects relevant to surface segregation in HEAs and providing advanced sampling capabilities of the HEA chemical and structural space without being specifically trained on HEA data. 

\begin{figure}
    \centering\includegraphics{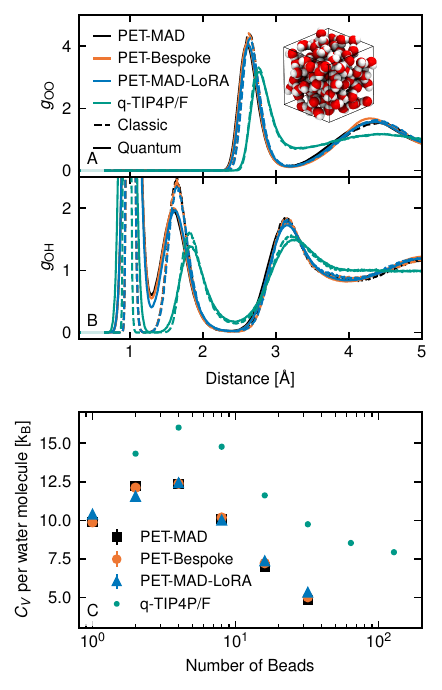}
    \caption{\label{fig:water-heat-capacity}
    Radial distribution functions $g$ of O-O and O-H pairs (top) and heat capacity $C_V$ of liquid water at 298 K and 1 atm, calculated using PET-MAD, a finetuned version of PET-MAD, and a bespoke PET model, as a function of the number of beads employed in the PIMD simulation (bottom). Results on the heat capacity obtained with the empirical q-TIP4P/F forcefield (reproduced from Ref.~\citenum{ceriotti2011accelerating}) are also included for comparison.}
\end{figure}

\subsection{Quantum  nuclear effects in liquid water}\label{ssec:liquid-water-results}

Water is a widely studied system because of its high biological, environmental and technological significance, as well as the many anomalies in its physical properties, which can be traced to the strong hydrogen bonds that form local tetrahedral motifs.
Due to its light hydrogen atoms, accounting for nuclear quantum effects (NQEs) is fundamental in order to extract accurate observables for liquid water, even at room temperature \cite{markland2018nuclear}. In this example, we employ path integral molecular dynamics (PIMD) to model nuclear quantum effects in the simulation of a medium-sized periodic system consisting of 128 water molecules (384 atoms).

We first evaluate the O-O and O-H pair correlation functions, which report on inter-molecular structural correlations, and show the extent of intra-molecular quantum fluctuations, respectively. Additionally, we calculate the constant-volume heat capacity at 298 K --- a physical quantity that exhibits very pronounced quantum effects, and that requires the use of sophisticated path integral estimators that are notoriously difficult to converge \cite{yamamoto2005path}. 
We compare (1) the PET-MAD model, (2) a system-specific PET model trained on the dataset from Ref.~\citenum{chen+19pnas}, with energies and forces recomputed at the same level of theory used for MAD, (3) a PET-MAD model which was finetuned on the same recomputed water dataset and (4) values for a simple classical forcefield that reproduces well most of the experimental properties of water and allows us to demonstrate the slow convergence of some of the estimators \cite{habe+09jcp}. 

The main takeaways from this comparison (shown in~Figure \ref{fig:water-heat-capacity}) are that (1) PET-MAD inherits the tendency of GGA functionals to overestimate the melting point of water, leading to room temperature simulations of a highly undercooled liquid, that is overstructured, with excessive delocalization of protons along H bonds, and a heat capacity that is closer to that of ice than to water; and (2) that the general purpose model is in excellent agreement with the bespoke models, both in classical simulations and in path integral simulations that are sufficient to converge the pair-correlation function, and approach convergence of the heat capacity. 
This is significant as it proves that PET-MAD can be used reliably even when probing the large intra-molecular distortions induced by zero-point fluctuations, and that it is fast enough to afford the large overhead of path-integral simulations with complex estimators -- even though it cannot avoid the limitations of the electronic-structure reference.

\subsection{Quantum nuclear effects in NMR crystallography}

We now consider the impact of quantum nuclear fluctuations in a different context, evaluating how they affect the nuclear magnetic resonance (NMR) chemical shieldings in organic crystals, and hence how they affect NMR crystallography.
In the prototypical NMR crystallography workflow, chemical shieldings are computed for a set of static candidate structures, and compared with those measured experimentally for an unknown polymorph \cite{hodgkinsonNMRCrystallographyMolecular2020a}. The most likely candidate structure is then taken to be the one for which the experimentally measured and computed shieldings best agree.

\begin{figure}
    \centering\includegraphics{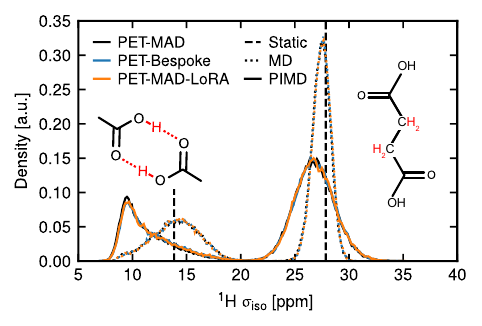}
    \caption{\label{fig:Shielding_distribution}
    Computed chemical shielding distribution of $^{1}$H in $\alpha$ succinic acid crystals, using PET-MAD (black),  PET-MAD-LoRA (orange) and PET-Bespoke (blue) models for PIMD (solid) and MD (dotted) simulations. Shieldings of the static geometry  (PET-MAD only) are shown as vertical lines (carboxylic H3 site around 15 ppm and an  average of the aliphatic H1 and H2 sites around 28 ppm). } 
\end{figure}

NMR shieldings result from the averaging of instantaneous values of structures distorted by thermal and quantum fluctuations, which can be sampled computationally by performing MLIP-driven (PI)MD simulations and then evaluating the chemical shieldings using a bespoke machine learning model for succinic acid crystals.
We simulate in particular the $\alpha$ polymorph of the succinic acid, one of the structures investigated in a previous study \cite{enge+21jpcl} from which we also obtain the training structures to build bespoke shielding and MLIP models (the test set accuracies (MAE) on energies\textbar{}forces are 12.5  meV/atom\textbar106.1 meV/Å  for PET-MAD, 3.1 meV/atom\textbar86.0 meV/Å for PET-Bespoke and 2.0 meV/atom\textbar64.5 meV/Å for PET-MAD-LoRA).
In line with the excellent accuracy of PET-MAD for molecular crystals, we observe near-perfect agreement between the distributions of chemical shieldings obtained with the three potentials (Figure \ref{fig:Shielding_distribution}), with both classical and quantum sampling -- the latter showing a large downward shift and broadening of the $^1$H shielding distributions, especially for the H-bonded protons, qualitatively the same observation that Engel and coworkers make in Ref.~\citenum{enge+21jpcl} who employ a bespoke MLIP trained on PBE0-MBD reference calculations. 
As in the case of water, these results demonstrate the ability of PET-MAD to describe quantum nuclear fluctuations. By combining MLIP-driven sampling with models of functional properties, such as the chemical shieldings, one can extend the reach of a general-purpose model beyond structural and energetic predictions.

\subsection{Dielectric response of barium titanate}\label{sec:BTO_result}

\begin{figure}
    \centering\includegraphics{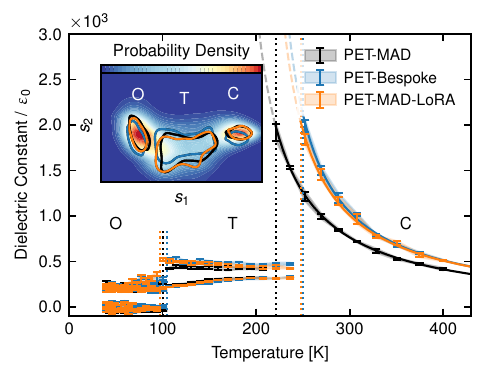}
    \caption{\label{fig:BTO_dielectric_constants}
    Temperature-dependent dielectric response of BTO for three PET models: the pre-trained PET-MAD (black), the bespoke PET model (blue) and a LoRA-finetuned PET-MAD (orange). 
    Vertical dotted lines correspond to phase transition temperatures estimated from the relative chemical potentials.
    The inset shows the collective variable landscape distinguishing three separate phases. The models consistently identify the different phases, as evidenced by the contour plots at 0.5 probability centered on the basins corresponding to each phase.
    }
\end{figure}

To conclude our benchmarking series, we study the ability of PET-MAD to describe the dielectric properties of materials using barium titanate system as an example. Barium titanate (BaTiO\textsubscript{3}, BTO) is a prototypical ferroelectric perovskite that undergoes a sequence of temperature-dependent structural phase transitions: rhombohedral (R, below 183\,K), orthorhombic (O, 183--278\,K), tetragonal (T, 278--393\,K), and cubic (C, above 393\,K)~\cite{merz1949electric}. The ferroelectricity in BTO arises from off-center displacements of the Ti atom within the oxygen octahedron, which break centrosymmetry in the lower-temperature phases. 
These displacements are fundamental to its ferroelectric behavior and phase transitions,\cite{bersuker1966origin} making BTO an ideal model system for studying ferroelectricity in perovskites.

To evaluate the capability of PET-MAD in characterizing ferroelectricity in this material, we perform flexible-cell MD simulations for a 320-atom model of BTO over a temperature range of 40--400\,K at constant ambient pressure, following the same protocol as in Ref.~\citenum{gigl+22npjcm}. Once again, we recompute consistent energetics for the material-specific dataset, to build bespoke and finetuned PET models. Test-set MAEs on energy{\textbar}forces are  12.67~meV/atom~\textbar27.96~meV/Å for PET-MAD, 0.23~meV/atom~\textbar9.41~meV/Å for the bespoke model, and 0.12~meV/atom~\textbar3.92~meV/Å for the LoRA-finetuned model, respectively. 

We then analyze the sampled configuration within a reduced-dimensionality collective variable landscape derived from a subset of neighbor density coefficients~\cite{gigl+22npjcm} (shown in the inset of~Figure \ref{fig:BTO_dielectric_constants}).
Measuring the populations of different phases by a clustering algorithm allows us to estimate their relative chemical potential, and hence the transition temperatures between the different phases.
Similar to what is observed in Ref.~\citenum{gigl+22npjcm}, the transition temperatures are heavily underestimated with respect to experiments (due to a combination of DFT shortcomings and finite-size effects), but there is excellent agreement between PET-MAD and the two bespoke models. The largest discrepancy, for the T-C transition, is below {30\,K}---a remarkable accuracy for a model with such broad applicability.

To further assess the reliability of PET-MAD in predicting functional properties, we also compute the temperature-dependent static dielectric tensor $\boldsymbol{\epsilon}_r$ (Figure \ref{fig:BTO_dielectric_constants}), by estimating the covariance of the total polarization, computed in turn using an equivariant linear model trained on the polarization dataset of Ref.~\citenum{gigl+22npjcm}.
All models predict qualitatively the large value of $\boldsymbol{\epsilon}_r$ for the paraelectric, cubic phase, which increases greatly as it approaches the ferroelectric transition temperature, and then takes smaller values in the tetragonal and orthorhombic phases (which have multiple inequivalent optical axes). 
As it is the case of the transition temperatures, the two specialized models are in near-perfect agreement, while the PET-MAD simulations slightly underestimate $\boldsymbol{\epsilon}_r$ in the T phase, and yield a curve in the C phase that is shifted by about 25\,K, consistent with the lower Curie temperature.
Overall, this final example demonstrates once more that, despite considerably larger test errors than for a bespoke model, PET-MAD captures with close-to-quantitative accuracy subtle physical effects, and can be used to drive complicated, advanced materials modeling protocols.

\section{Discussion}

PET-MAD pushes the boundaries of what generally-applicable ``universal'' machine learning potentials can achieve, providing out-of-the-box semi-quantitative accuracy for both inorganic and organic materials benchmarks, as well as for six demonstrative applications covering a very broad range of material types and advanced simulation techniques.
Moreover, it does so while being trained on a data set that is orders of magnitudes smaller than those used for current state-of-the-art models. This observation suggests that the computational budget for training set construction can be best allocated by pursuing internal consistency and a high degree of structural and chemical diversity --- two guiding principles which underlie the construction of the MAD dataset we introduce here.
A compact dataset facilitates model training and optimization, as well as the transition to more accurate, and computationally demanding, approximations to the electronic-structure problem. 

The PET architecture we use is not constrained to follow rigorous rotational invariance; nonetheless, it learns to make predictions that are invariant to a high degree of accuracy. It achieves a high expressiveness thanks to its transformer module, while being faster than all the conservative models we considered, and demonstrating absolute speed records in the non-conservative regime. 
The availability of a reliable and inexpensive uncertainty quantification framework makes it possible to assess model error, and also its propagation to the ultimate quantity of interest. 
In cases where this error is larger than desired, a simple fine-tuning procedure can be applied to further improve the accuracy for a specific target system, while maintaining the ability to make qualitative predictions for the high-diversity test set.
PET-MAD is easily accessible from several atomic-scale simulators thanks to its integration within the \texttt{metatensor}\cite{METATRAIN} ecosystem. 
It provides an efficient and accurate solution to perform simple and advanced atomistic simulations of materials, and we expect it will serve as inspiration for future developments of both datasets and model architectures, as well as a reliable engine to drive the design and discovery of new and better materials.

\section{Methods}

We provide a brief but complete overview of the procedure we followed to construct the MAD dataset, to train the PET-MAD model, and to fine-tune single-purpose models. 
The simulation protocols for the advanced simulation examples (Sections ~\ref{sec:LPS_result}-\ref{sec:BTO_result}) follow closely those of the cited publications, and are briefly discussed in the SI for the convenience of the reader. 

\subsection{MAD dataset construction }\label{sec:mad_details}

In four of the MAD subsets, namely MC3D\cite{Huber2022MC3D}, MC2D\cite{Campi2023MC2D}, SHIFTML-molcrys\cite{Cordova2022} and SHIFTML-molfrags\cite{Cersonsky2023}, we simply used the  published structures, recalculating their energy and forces with the MAD DFT settings provided in Section \ref{sec:dft_details}. For the remaining MC3D-based subsets, we always started from the original MC3D crystals and made different structural modifications, with the sole goal of improving coverage of the configuration space for highly-distorted, unusual configurations.
Structures in the MC3D-rattled subset were obtained by selecting each MC3D crystal with more than one atom in the unit cell and applying to the Cartesian coordinates of each atom a Gaussian noise with zero mean and a standard deviation equal to 20\% of the corresponding covalent radii. 
The MC3D-randcomp subset was created by first taking the atomic positions and cell parameters of a random crystal structure from the MC3D subset, then assigning random atom types to every lattice site, and finally adjusting isotropically the total volume of the cell to match the total atomic volume calculated based on the covalent radii of the elements. 
The surface slabs in the MC3D-surface subset were obtained by cleaving a randomly-selected MC3D crystal along a random, symmetrically distinct crystallographic plane with the maximum value of the Miller index (hkl) equal to 3, also ensuring the orthogonality of the normal lattice vector to the surface plane.
The MC3D-cluster subset was constructed by cutting a random spherical atomic environment containing between 2 and 8 atoms from a randomly selected MC3D crystal structure.

\subsection{First-principles calculations}\label{sec:dft_details}

All datasets in this work (i.e. MAD and the reference subsets used for benchmarking, as well as the single-purpose dataset for the six case studies) were built by computing energies and forces using an identical  density functional theory (DFT) setup. We used the Quantum Espresso v7.2 package \cite{gian+09jpcm}, compiled with the SIRIUS\cite{SIRIUS} libraries as the primary DFT engine and AiiDA \cite{Huber2020} as a workflow and task manager. 
To reduce the possibility that some simulations would converge to ambiguous magnetization states, all simulations were performed in a non-magnetic  setting, using the PBEsol exchange-correlation functional. 
Even though this choice is known to have shortcomings for several classes of materials -- as shown by the discrepancy of some of our examples with experiments -- it allows us to generate predictions for a highly diverse system with a high degree of stability and converged numerical parameters. 
We described valence and semi-core electrons using the standard solid-state pseudopotentials library (SSSP) v1.2 (efficiency set) \cite{pran+18npjcm}, with the most stringent settings for plane-wave and charge-density cutoffs across the 85 elements (110 Ry and 1320 Ry, respectively).
As discussed in the SI, using different settings to match the maximum cutoff of each structure would give substantial inconsistencies between the description of some atomic types. 
Electronic smearing and partial occupancies were described with a Marzari-Vanderbilt-DeVita-Payne cold smearing 
with a spread of 0.01 Ry. In all periodic dimensions, the first Brillouin zone was sampled with a $\Gamma$-centered grid with a resolution of 0.125 Å$^{-1}$, while only a $\Gamma$ point was used along the non-periodic dimensions. 
In the case of non-periodic systems we also applied a truncation scheme of the Coulomb potential to avoid the interaction of periodic replicas of the system through the periodic boundary conditions: the Sohier-Calandra-Mauri method 
for 2D systems and the Martyna-Tuckerman correction
for 0D systems. 
Along each non-periodic direction we additionally applied a 25~Å vacuum to ensure convergence of the truncation methods. A compositional baseline based on isolated atom energies was subtracted from the DFT energies of each structure to improve the numerical stability of training. 

For most of the MAD subsets listed in Section \ref{sec:dataset_details}, the DFT settings provided above ensured a good convergence rate of $>95\%$, i.e. convergence was not achieved in less than 5\% of the cases. The exception is MC3D-random, for which only about 55\% of the attempted simulations converged.
This result does not come as a surprise and can be explained by the extremely non-equilibrium state of these structures.

\subsection{PET model}\label{sec:pet_details}

As our machine learning approach for fitting interatomic potentials, we employ the \emph{Point Edge Transformer} (PET)~\cite{pozd-ceri23nips}, which has been shown to achieve state-of-the-art performance across a diverse set of benchmarks spanning both molecular and materials domains. 
In essence, PET is a graph neural network (GNN) where each message-passing layer is implemented as an arbitrarily deep transformer. 
Specifically, PET maintains feature vectors (or \emph{messages}) $f_{ij}^l$ for every directed bond between atoms $i$ and $j$ that lie within a specified cutoff radius. Here, the superscript $l$ denotes the message-passing layer index, and each $f_{ij}^l$ is a fixed-size vector (a token) of dimension $d_{\mathrm{PET}}$. These intermediate representations are updated at each message-passing layer by  a transformer with an adjustable number of internal layers, invoked for each atomic environment.
At each message-passing layer, for each atom $i$, the input tokens to the transformer consist of the message vectors $\{ f_{ji}^l \}_j$ coming from all neighbors $j$ to the central atom $i$. The transformer then applies a \emph{sequence-to-sequence} transformation in a permutation-covariant manner. Its outputs are subsequently interpreted as the new set of outbound messages from atom $i$ to each neighbor $j$, $\{ f_{ij}^{l+1} \}_j$. Geometric information (i.e., the 3D positions of neighbors) and chemical species are also incorporated into this process. 
To obtain the desired target property (e.g., energy), PET applies feed-forward neural networks to each representation $f_{ij}^l$ and sums all the outputs across bonds $ij$ and layers $l$.
The PET architecture imposes no explicit rotational symmetry constraints, but learns to be equivariant through data augmentation. This unconstrained approach yields high theoretical expressivity: even a single layer of the model acts as a universal approximator featuring virtually unlimited body order and angular resolution.
A detailed specification of the architecture and the full functional form can be found in Ref.~\citenum{pozd-ceri23nips}.

\subsection{Training of PET-MAD}\label{sec:pet-mad_details}

We split the MAD dataset into training, validation, and test parts by separately shuffling each of the eight subsets, and selecting respectively 80\%, 10\%, and 10\% of the structures. 
The composition contribution to the total energy of the structures was fitted using a simple linear model and subtracted to improve the training behavior. 
We used a Pareto-optimal architecture (see the details on hyperparameters optimization in the Supplementary Text) with a cutoff radius of 4.5 Å, 2 message-passing layers, each containing 2 transformer layers with a token size of 256, 8 heads in the multi-head attention layer, 512 neurons in the output multi-layer perceptron. Training was performed using the PyTorch framework and the \texttt{metatrain} package \cite{METATRAIN} on 8 NVIDIA H100 GPUs with a batch size of 24  structures per GPU for a total of 1500 epochs and took about 40 hours. 
We used the Adam 
optimizer with an initial learning rate (LR) of $10^{-4}$ and applied a LR scheduler, which halved the LR every 250 epochs. The loss function was based on a root-mean-square error difference between the model's energy and force predictions and the corresponding target values, with a scaling factor of 0.1 applied to the energy contribution. Mean absolute errors of the trained PET-MAD model were 7.3 meV/atom\textbar43.2 meV/Å \ in energies and forces prediction on a train set. Validation set errors were 14.7 meV/atom\textbar72.2 meV/Å \ , respectively. 

\subsection{Fine-tuning}\label{sec:fine-tuning-details}

Fine-tuning of the pre-trained PET-MAD model was performed using a parameter-efficient fine-tuning technique based on low-rank adaption (LoRA) ~\cite{LoRA}. This method is widely used in the ML community, since it allows to mitigate the \textit{catastrophic forgetting} --- the phenomenon of losing the accuracy on the base dataset while tuning the model on the new dataset, which is inherent to the conventional fine-tuning, where all the weights of the model are trainable~\cite{Biderman2024}.  
In the LoRA approach, the weights of the base model are frozen during the training, while an additional set of trainable weights composed of two low-rank matrices is added to each attention block of the model with a regularization factor, which controls the influence of the low-rank matrices on the model’s weights. By adjusting rank and regularization factor, one can optimize the fine-tuning workflow to achieve better performance on specific tasks, without compromising entirely the ability to perform the more general task. In this work, we used a rank of 8 and a scaling parameter of 0.5 for all LoRA finetuned models, unless otherwise specified.

In terms of raw accuracy on the specialized datasets, the results in~Section \ref{sec:results} and~Section \ref{sec:learning-curves} indicate that fine-tuning is always beneficial compared to training a specialized ad-hoc model in the low-data regime. On larger datasets, while in some cases a specialized model trained from scratch \textit{can} exceed the accuracy of a finetuned model, this is not the case for many of the systems investigated in this work, including barium titanate, succinic acid, lithium thiophosphate, and high-entropy alloys.

LoRA-finetuned models retain varying degree of accuracy (see the Section \ref{sec:fine-tuning-accuracies} for details) on the generic structures from the MAD dataset, while still providing practically equivalent observables from an ad-hoc fully specialized model. In general, we therefore recommend to employ LoRA if finetuning PET-MAD for a specific application.

\subsection{Uncertainty quantification}~\label{sec:uq}

We provide uncertainty quantification for the PET-MAD model via the LLPR method~\cite{bigi+24mlst}, which produces \textit{a posteriori} uncertainty estimates for trained neural networks. 
The LLPR uncertainties can be computed based on the covariance of the last-layer features over the training set, and they can be evaluated at nearly no additional cost compared to the raw predictions. The LLPR formalism is particularly suitable in the context of this study, as it can be used to sample a finite number of last-layer weight sets. The resulting last-layer ensemble can be used to propagate uncertainties through arbitrarily complex atomistic workflows, by manipulating separately the outputs of the different ensemble members, and/or using them to reweight the trajectories generated by the mean of the ensemble~\cite{imba+21jcp}.

\begin{acknowledgments}
The Authors would like to thank Giovanni Pizzi, Marnik Bercx and Sebastian Huber for helping with the set-up of AiiDA calculations on the MC3D dataset, and Sandip De for discussion on potentials for high-entropy alloys.
They are also grateful to all the current and past members of the Laboratory of Computational Science and Modeling who contributed to the software infrastructure that supported this work.
Computation for this work relied on resources from the Swiss National Supercomputing Centre (CSCS) under the projects s1092, s1219, s1243, s1287, lp26,  Finnish IT Center for Science (CSC) under the project 465000551, and the EPFL HPC platform (SCITAS). 

\paragraph*{Funding:}

AM and MC acknowledge support from an Industrial Grant from BASF and from EPFL.
GF and MC acknowledge funding from the MARVEL National Centre of Competence in Research (NCCR), funded by the Swiss National Science Foundation (SNSF, grant number 182892)
SP, FB and GF Platform were supported by a project within the Platform for Advanced Scientific Computing (PASC).
DT acknowledges support from a Sinergia grant of the Swiss National Science Foundation (grant ID CRSII5\_202296).
PL, PP and MC acknowledge the funding from the European Research Council (ERC) under the European Union’s Horizon 2020 research and innovation programme (grant agreement No 101001890-FIAMMA).
MK and MC acknowledge support by a Swiss National Science Foundation (grant ID 200020\_214879).

\paragraph*{Author contributions:}
A.M. worked on creation of the MAD dataset, training the PET-MAD model, running the accuracy benchmarks for PET-MAD and reference models, performing the calculations of surface segregation in CoCrFeMnNi. F.B. worked on implementation of the \texttt{metatrain} infrastructure for training and evaluating the PET-MAD model, as well as performing the uncertainty quantification for it, performed the speed benchmarking for PET-MAD and reference models, and studied the nuclear quantum effects in water. M.K. calculated the melting point in GaAs and studied the chemical shielding in succinic acid. P.P. studied the dielectric response in barium titanate. D.T. studied the ionic conductivity in lithium triophosphate. G.F. developed the \texttt{metatensor} ecosystem, which \texttt{metatrain} infrastructure is based on, developed the interfaces for ASE and LAMMPS simulation engines, and helped with the computational setup for creating the MAD dataset. S.P. developed the original version of PET architecture. P.L. worked on implementation of the \texttt{metatrain} infrastructure and finalized the figures. M.C. designed and guided the project, and provided a theoretical support. All authors contributed to the original version of the manuscript. 
\paragraph*{Competing interests:}
There are no competing interests to declare.

\paragraph*{Data and materials availability:}

The accompanying data sets and simulation input files will be available in public repositories, such as the Materials Cloud Archive. Pre-trained PET-MAD model, environments providing all necessary dependencies, will be made available upon publication.
All code contributions are available as public \emph{git} repositories, with simplified installation instructions available at \url{https://github.com/lab-cosmo/pet-mad}.
A practical usage example can be found at \url{http://atomistic-cookbook.org/examples/pet-mad/pet-mad.html}.

\end{acknowledgments}

\setcounter{enumiv}{0}

\clearpage

\renewcommand{\thefigure}{S\arabic{figure}}
\setcounter{figure}{0}
\renewcommand{\thesection}{S\arabic{section}}
\setcounter{section}{0}

\onecolumngrid
\part*{Supplementary Information}
\twocolumngrid

\section{Hyperparameters optimization}

To obtain the optimal model in terms of accuracy and computational speed, we performed a grid search over the most important hyperparameters that define the architecture of the model. The hyperparameters and the values they were ranged in are provided below:

\begin{enumerate}
    \item The cutoff radius of the model ($R_\mathrm{cut}$) - [4.0, 4.5, 5.0, 5.5]
    \item Number of message-passing (MP) layers ($N_\mathrm{GNN}$) - [1, 2, 3, 4]
    \item Number of transformer layers in each MP layer ($N_\mathrm{trans}$) - [1, 2, 3, 4]
    \item Dimensionality of the hidden space ($d_\mathrm{PET}$) - [64, 128, 256]
    \item Number of heads in the multi-head attention layers ($N_\mathrm{heads}$) - [4, 8]
\end{enumerate}

\begin{figure}
    \centering
    \centering\includegraphics{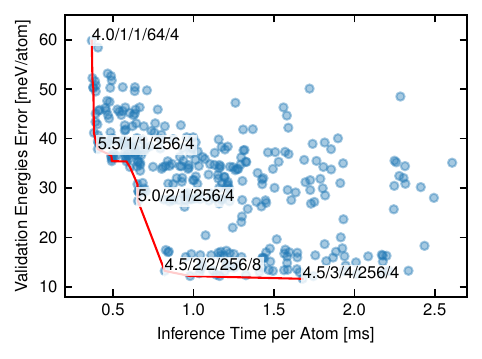}
    \caption{Pareto frontier of the PET-MAD models performance with different architectures. Each architecture is represented as a single point on a scatterplot. Inference time per atom on a single NVIDIA GH200 GPU with a batch size of 1 is plotted along the x-axis. The mean absolute error (MAE) in predicting energies on the validation set is plotted along the y-axis. The Pareto frontier is drawn as a solid red line. A few selected architectures are highlighted using the following notation: $R_\mathrm{cut}/N_\mathrm{GNN}/N_\mathrm{trans}/d_\mathrm{PET}/N_\mathrm{heads}$. The optimal model has hyperparameters of 4.5/2/2/256/8. } 
    \label{fig:pareto-frontier}
\end{figure}

For each combination of the hyperparameters a separate training was performed, and model accuracy and inference time on a validation set was evaluated thereafter using a single NVIDIA GH200 GPU with a batch size of 1 -- which is representative of the performance of the model for molecular dynamics. The resulting data was plotted on a single scatter plot (Figure \ref{fig:pareto-frontier}), and the Pareto-frontier construction was used to find the optimal set of hyperparameters. Based on this analysis, we conclude that the optimal PET-MAD model has a cutoff radius of 4.5 \AA, 2 MP layers with 2 transformer layers each, 256 neurons in the hidden space and 8 heads in the multi-head attention layers. 

\section{Details of benchmarking subsets selection}

We compared the accuracy of PET-MAD against four recent universal machine learning interatomic potentials - MACE-MP-0 L, MatterSim-5M, Orb-v2, and SevenNet-l3i5 - on popular atomistic ML datasets: MPtrj, Matbench Discovery, Alexandria, SPICE, MD22, and OC2020 S2EF. For each dataset (including MAD), we prepared a small subset of structures that were recalculated using DFT settings consistent with each model's training data.

The MAD benchmark consists of 360 structures selected by sampling 50 random structures from each MAD test subset, recalculated with MPtrj DFT settings, and cleaned of non-converged structures and outliers. The MPtrj benchmark is based on the MACE-MP-0 validation subset, reduced to 136 structures after removing four 1D wire structures. The Matbench Discovery benchmark contains 555 structures, randomly sampled from the original WBM dataset \cite{WBM-dataset} - a part of the Matbench Discovery, which addresses the structural stability of inorganic crystals with random elemental substitutions. We didn't include the structures with lanthanides and actinides to this subset in order to balance the effect of low coverage of these elements in the MAD dataset compared to other datasets like MPtrj.
The OC2020-S2EF benchmark consists of 78 structures, where 100 structures were first sampled from the OC2020-S2EF training dataset and then cleaned of non-converged cases. 
The SPICE benchmark contains 99 structures, where 100 structures were randomly sampled neutral molecules from the SPICE dataset, and later cleaned of non-converges ones. The Alexandria benchmark includes 200 structures: 50 were randomly sampled from Alexandria-2D and Alexandria-3D-gopt, and 100 from Alexandria-3D. The MD22 benchmark consists of 134 structures, where 25 structures were first randomly sampled from each of the seven subsets of the original MD22 dataset (Ac-Ala3-NHMe, AT-AT, DHA, Stachyose, AT-AT-CG-CG, Buckyball-Catcher, double-walled-nanotube), and then cleaned of non-converged cases.

\section{Assessing PET-MAD on Matbench Discovery benchmark} \label{sec:matbench-discovery}

\begin{figure}[tbh]
    \centering
    \includegraphics[width=\linewidth]{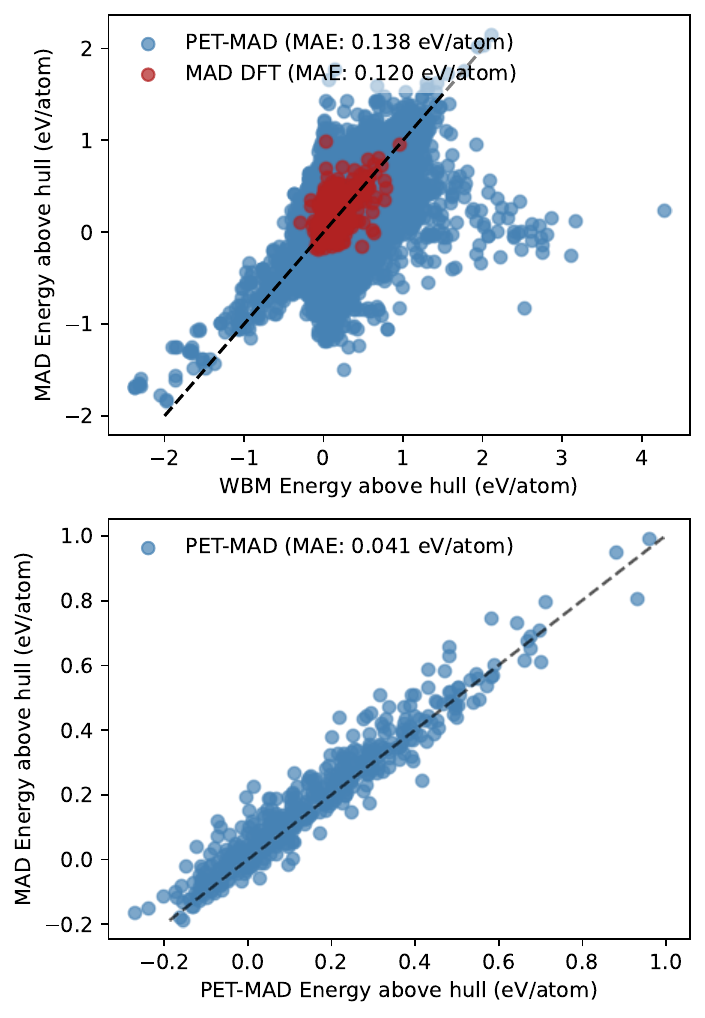}
    \caption{Right panel: comparison of the energies above hull of a subset of structures from the WBM dataset \cite{WBM-dataset}, recomputed with DFT using MAD settings (MAD DFT, red dots), and predicted with PET-MAD (blue dots). The mean absolute error (MAE) in predicting the WBM values of the energy above hull in eV/atom is presented in the legend. This comparison demonstrates the upper limit of accuracy of any MAD-trained on the Matbench-Discovery benchmark upon using non-consistent DFT settings, as the difference in baseline DFT energy between Matbench and MAD settings reaches 120 meV/atom. Left panel: comparison of the PET-MAD predictions of the energy above hull against the consistent reference, recomputed with MAD settings. The resulting error decreases from 138 to 41 meV/atom, revealing the actual accuracy of the model.}
    \label{fig:mad-wbm}
\end{figure}

As discussed throughout this work, it is crucial to maintain consistency in the level of DFT theory in the training set of the model. Lack of consistency in training data can introduce systematic errors to the model predictions, which can sometimes go up to 30 meV/atom (see Figure \ref{fig:energy_difference_mad_vs_defaults} for details). It is equally important to keep the consistency while assessing the accuracy on the model on benchmarks, like the popular Matbench Discovery \cite{Matbench}. Figure \ref{fig:mad-wbm} demonstrates the results of the PET-MAD assessment on the WBM dataset \cite{WBM-dataset} (a part of the Matbench Discovery) in predicting the energies above hull. When comparing the predictions against the default WBM values, PET-MAD yields a large error of around 140 meV/atom. However, this error is almost entirely explained by the significant discrepancy of 120 meV/atom in the underlying DFT reference, as demonstrated by comparing the WBM values to those recomputed using the MAD dataset settings on a subset of the WBM crystals. This value essentially sets an upper limit on the accuracy of any MAD-trained model when compared to data obtained with inconsistent (in this case - WBM) DFT settings. However, while using the consistent level of theory, one can significantly improve the models' accuracy: in the left panel of the Fig. \ref{fig:mad-wbm} we demonstrate the actual accuracy of the PET-MAD model if compared to the consistent DFT reference: the MAE drops from 138 meV/atom to 41 eV/atom.

This final MAE value was used to address another important property of the universal MLIPs - their data-efficiency. For all the models in Fig. \ref{fig:umlip_frontier} in the main text, we calculated the MAE value in predicting the energy above hull based on the same WBM subset, while utilizing the default WBM energy values for all the models except PET-MAD.

Therefore, since there is no straightforward way to account for significant discrepancies in the baseline DFT data in the Matbench Discovery benchmark, we decided not to add PET-MAD to the models list on the benchmark website. The resulting accuracy numbers would not be representative due to the lack of consistent DFT reference. We also didn't include the phonon properties related part of the Matbench Discovery benchmark in this discussion, as we instead performed a more systematic study of the phonon properties in Supplementary Section \ref{sec:phonons}.

\section{Effect of \textit{ab initio} calculations settings convergence}

\begin{figure*}
    \centering\includegraphics[width=\textwidth]{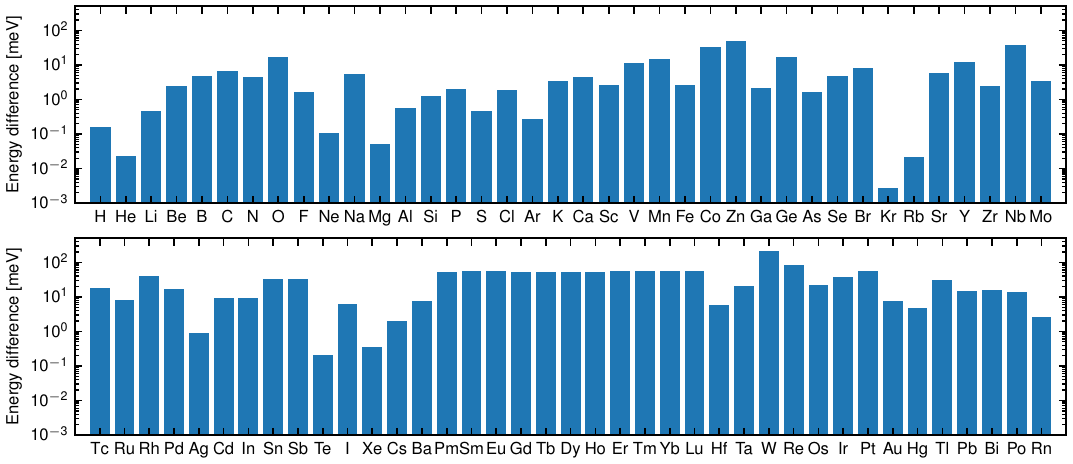}
    \caption{Absolute difference in energies of isolated atoms calculated with MAD dataset settings and recommend SSSP values of the plane waves basis set and charge density cutoffs.}
\label{fig:energy_difference_mad_vs_defaults}
\end{figure*}

One of the main features of the MAD dataset developed in this work is its high internal consistency in the DFT settings used. This implies the use of extremely high cutoff values for the wavefunction basis set (110 Ry) and charge density (1360 Ry), regardless of the considered system and set of elements. These values effectively are consistent with the most restrictive recommended settings that are provided in the SSSP pseudopotential library \cite{pran+18npjcm} (which range from 30 / 120 Ry to 90 / 1080 Ry for the wavefunctions basis set / charge density cutoffs, respectively)  for the elements included in the MAD data set (except for Radon, which has the recommended cutoffs of 120 Ry and 960 Ry, respectively). 
To explain the necessity for this protocol (rather than that one would use in first-principles calculations, that usually involves choosing the most restrictive settings \emph{within each structure}) we compare the energies of isolated atoms (which we use as a baseline in MAD dataset), calculated using MAD settings and the recommended SSSP settings for each element (Figure \ref{fig:energy_difference_mad_vs_defaults}). 
These results show that for certain elements the difference in energies between the recommended and MAD settings can reach 10-100 meV/atom, which is either comparable or higher than the typical error of PET-MAD. If the level of convergence depended on the composition of each structure, the energy contribution of each atom would not depend exclusively on the atoms within the receptive radius of the model, but on the global composition, which is unphysical and ultimately unlearnable. 
While it may be possible to carefully correct for these effects (that do not usually affect \emph{relative energies} within a fixed composition space) we prefer to avoid the inconsistency altogether, despite the substantial computational overhead in training-set construction.

\begin{figure*}
    \includegraphics[width=\textwidth]{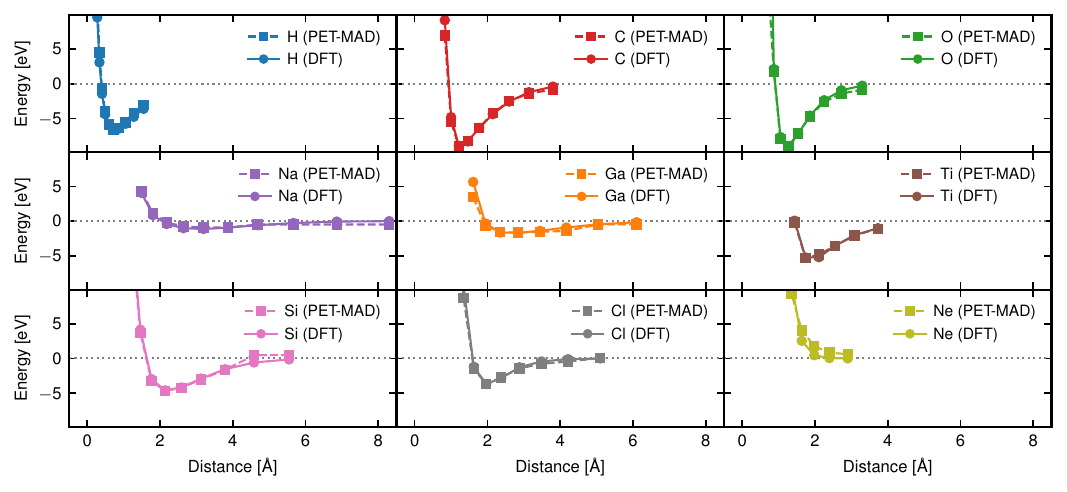}
    \caption{Energy dependence of elemental dimers on different interatomic distances for several elements, calculated with MAD DFT settings and the PET-MAD model. All dimers were treated as diatomic clusters and shared the DFT calculation protocol with the MC3D-cluster subset.  }
\label{fig:dimer_curves}
\end{figure*}

\section{Diatomic energy curves}

Lack of training data in the region of small interatomic distances can lead to a phenomenon of \textit{artificial dimers}, where a pair of atoms essentially merge into a single point in space due to no explicit repulsive part in the potential. One way to solve this problem is to include constraints in the model, 
which define the behavior at short interatomic distances. Another way is to make the model learn this behavior by adding the appropriate configurations with close atomic distances to the data set. In this work, we use a second approach while training the PET-MAD model, and it yields quantitatively accurate diatomic curves with correct repulsive behavior (Figure \ref{fig:dimer_curves}). This is mainly due to the MAD dataset, which is diverse enough to provide the necessary data in the range of short interatomic distances. The set of distances on which we evaluated the energies of dimers is based on a logarithmic grid of ten points ranging from 0.9 to 5.0 of the corresponding element's covalent radius in angstroms.

\section{Geometry optimization with PET-MAD}

\begin{figure}
    \centering
    \includegraphics[width=\linewidth]{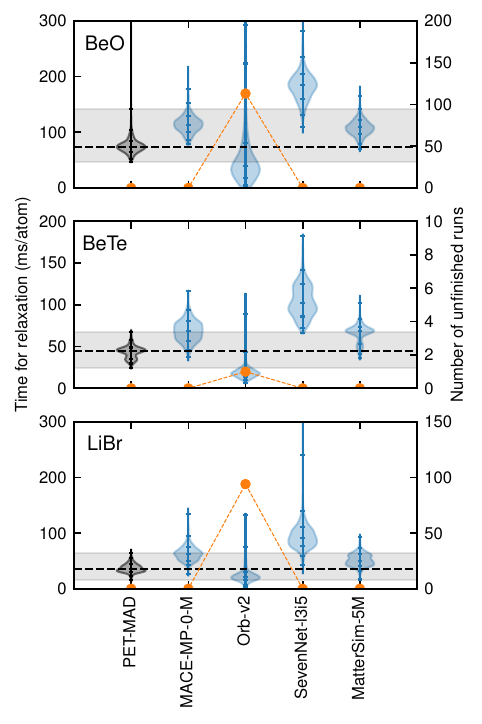}
    \caption{Timing distributions for 1000 geometry optimization runs using the L-BFGS optimizer. The black dashed band represents PET-MAD timings within three $\sigma$ from the median under the Gaussian approximation (0.135th to 99.865th percentile). The orange circles are the number of times the optimization runs did not complete.}
    \label{fig:geomopt}
\end{figure}

We evaluated PET-MAD for geometry optimization on the three materials from \autoref{sec:phonons}, namely, wurtzite BeO, zincblende BeTe, and rocksalt LiBr taken from the phononDB dataset. We used a simple optimization workflow using ASE calculators and the LBFGS optimizer with a default maximum force threshold of 0.05\,eV/\AA. To collect statistics, we performed 1000 optimization tasks per material, applying random displacements to atomic positions and lattice vectors from a normal distribution with zero mean and standard deviation of 0.1\,\AA. We measured the time to complete each optimization for the five universal MLIPs used in our analyses and show the time distributions in \autoref{fig:geomopt}. We limited the maximum number of optimization steps to 100. PET-MAD timings are comparable to or faster than those of all conservative models tested. The non-conservative Orb-v2 is consistently faster but less stable, being the only model that failed to converge in up to 10\% of the displacements.

\section{Uncertainties of PET-MAD}

\begin{figure*}
    \includegraphics[width=\textwidth]{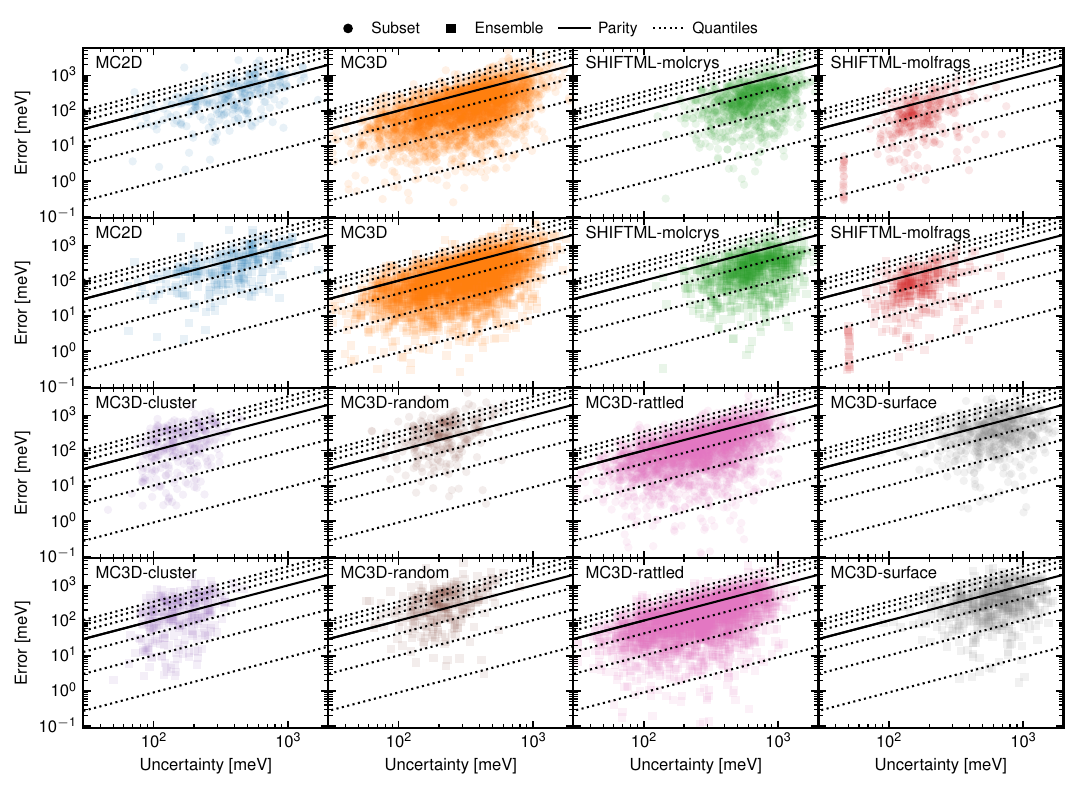}
    \caption{\label{fig:pet-mad-uncertainties}
    Upper panels per color: Predicted vs. actual per structure error of the PET-MAD model on the MAD test set within the LLPR protocol.
    Lower panels per color: same error but after converting the LLPR uncertainty model to a shallow ensemble}
\end{figure*}

Replacing expensive numerical computations with efficient statistical models is the centerpiece of this manuscript. Predictions of the universal force field are potentially uncertain with respect to the reference method it aims to approximate. 
Model uncertainties of universal force fields are dominated by epistemic contributions that arise from a lack of knowledge of the model about an area of the structural and compositional space that is only loosely covered by the training set. 
Given the vast compositional and structural space, any universal machine learning forcefield will suffer from areas of low coverage of training samples, especially since new, undiscovered classes of material could never have been considered in the construction of the training set. 
Epistemic uncertainties are by no means a reason why universal forcefields are bound to fail from the outset; rather, it is necessary to  manage them carefully, by equipping universal models with uncertainty quantification schemes, that indicate to the user when model predictions are uncertain and need to be treated with caution. Model errors also propagate to derived quantities, such as thermodynamic averages of observables.

Oftentimes model uncertainties in atomistic machine learning models are estimated using ensembles or committees of models, due to their conceptual simplicity and ease of implementation, requiring only to train a series of models and evaluating them. Albeit conceptually simple and easy to implement, full ensembling increases training and evaluation cost linearily with the number of models in the ensemble. The LLPR formalism~\cite{bigi+24mlst} offers an alternative way to obtain cheap ensembles from a trained neural network.
Within the LLPR method, predictive uncertainties are computed as
\begin{equation}
\sigma_i^2 = \alpha \, \mathbf{f}_i^\top (\mathbf{F}^\top \mathbf{F} + \varepsilon^2 \mathbf{I})^{-1} \mathbf{f}_i.
\end{equation}
In this expression, $\sigma_i$ is the uncertainty on the prediction relative to sample $i$, $\mathbf{f}_i$ are the latent features in the last layer of the neural network, $\mathbf{F}$ is a matrix whose rows correspond to each set of last-layer features for each structure in the training set, and $\varepsilon$ is a small regularizer. 
Besides computing energy uncertainties at nearly no additional cost compared to the raw predictions, the LLPR also allows the sampling of a last-layer ensemble~\cite{bigi+24mlst}, whose uncertainties can be propagated numerically through arbitrarily difficult workflows, including molecular dynamics~\cite{kell-ceri24mlst}. The GaAs calculations in~Section \ref{sec:GaAs_result} demonstrate how combining LLPR-sampled last-layer ensemble models with thermodynamic reweighting enables uncertainty quantification of PET-MAD predictions and propagate their uncertainties to thermodynamic averages at virtually no additional cost, at both training and inference time. The LLPR uncertainties, as well as those from an LLPR-derived last-layer ensemble of 128 members, are shown in Figure \ref{fig:pet-mad-uncertainties} for energy predictions on the MAD test set. The predictions follow the expected distribution almost exactly (see, for example, Refs.~\cite{bigi+24mlst} and~\cite{kell-ceri24mlst} for a discussion of this type of plots), confirming the very high quality of PET-MAD uncertainties, at least within the training domain.

\section{Uncertainty estimation for phonon dispersion curves}\label{sec:phonons}

\begin{figure*}[htb]
    \centering
    \includegraphics[width=\linewidth]{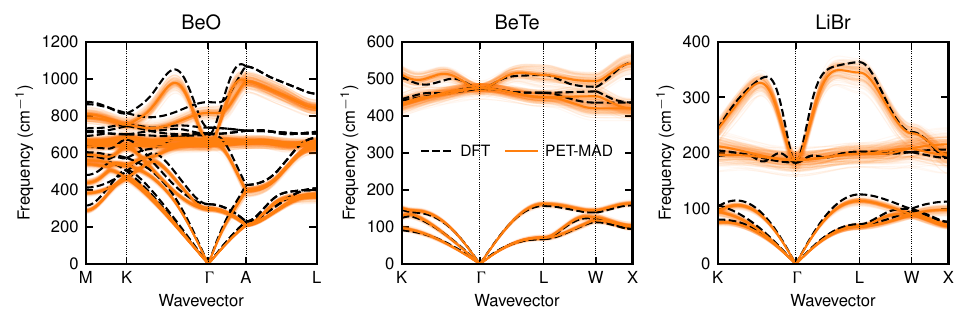}
    \caption{LLPR ensembles of phonon bands (orange, thin lines) for three representative structures of the phononDB dataset, compared with reference DFT results (black, dashed lines). Average values of the bands are reported denoted by orange, thick lines.}
    \label{fig:phonons}
\end{figure*}

In addition to complex properties that require MD sampling, such as the melting temperature discussed in the main text, uncertainty can also be estimated for static properties like phonon band structures. In Fig.~\ref{fig:phonons}, we present the LLPR ensemble of phonon bands for three representative materials from the PBEsol phononDB dataset~\cite{phonondb1}: wurtzite BeO, zincblende BeTe, and rocksalt LiBr, which were also analyzed in Ref.~\citenum{pota2024thermal} using the MACE-MP-0 foundation model. 

In the case of BeO, the optical bands appear slightly softer than in the reference calculations, which we checked to produce phonon bands consistent with calculations done with MAD settings with a frequency root mean square deviation of 3\,cm$^{-1}$. This behavior, however, is commonly observed for universal MLIPs, as shown in Ref. \cite{loew2024universal}, and occurs for BeO as well as for the other materials presented here when using MACE-MP-0~\cite{pota2024thermal}. 
The deviation is reflected in the UQ, which shows increased variance in the phonon band ensemble for optical modes.

\section{Fine-tuning accuracies}\label{sec:fine-tuning-accuracies}

For each simulation case presented in this work we trained a bespoke PET model from scratch, and compared it against the LoRA-finetuned version. While being equally accurate in predicting observables, the fine-tuned model retains a certain degree of accuracy on the base MAD dataset, which can be beneficial in certain computational setups. In Table \ref{tab:fine-tuning-models-mad-accuracies}, we list the mean absolute errors of each fine-tuned model in predicting the energies and forces on the base MAD test set. For reference, the general-purpose PET-MAD model yields errors of 15.1 meV/at.~\textbar~72.3 meV/\AA{} on the MAD test set.

\begin{table}[btp]
    \caption{\label{tab:fine-tuning-models-mad-accuracies}
    Accuracies of the LoRA-finetuned models in predicting the energies (in meV/at.) and forces (in meV/\AA) on the MAD test set. For the pre-trained PET-MAD errors are 15.1 meV/at.~\textbar~72.3 meV/\AA.}
    \centering\begin{tabular}{lcc} \toprule 
    LoRA Model &  Energy MAE &  Forces MAE\\ 
    & meV/at. & meV/\AA \\
    \midrule 
    LPS& 129.4& 215.1\\
    GaAs& 78.8 &  134.5\\
    HEA25S& 91.1& 228.9\\
    Water& 284.8&  288.3\\
    BTO& 44.4& 140.8\\
    Succinic acid& 144.4& 191.1\\
   \bottomrule
\end{tabular}
\end{table}

\section{Learning curves}\label{sec:learning-curves}

\subsection{PET-MAD}

\begin{figure}
    \centering
    \includegraphics{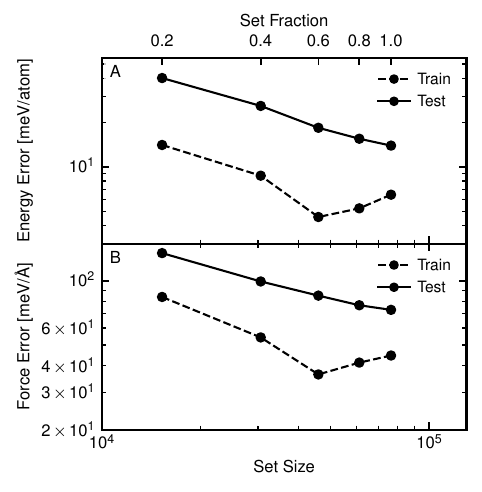}
    \caption{Learning curves of the PET-MAD model. The random fraction of the full MAD training set used for training is plotted on a x-axis, and the mean absolute error in predicting energies and forces on a MAD test set is given on a y-axis.}
    \label{fig:pet-mad-lc}
\end{figure}

Figure \ref{fig:pet-mad-lc} shows the learning curves of the PET-MAD model. Each training was performed on a fraction of MAD dataset, where a corresponding fraction of the structures was sampled from each subset, and then unified and shuffled. Even after training on only 20 \% of the data, the model achieves a reasonable accuracy with a mean absolute error in predicting forces of the MAD test set of 134.7 meV/\AA. With further increase in the training dataset size, the test MAE for both energies and forces gradually reduces. After reaching 60\% of the training data, we observe that the errors on a train set start growing up, which means that the dataset is diverse enough that the model cannot overfit. This saturation is not an indication of intrinsic limitations of the PET architecture, but of the fact that - for the small size of the MAD dataset - the Pareto optimal architecture corresponds to a comparatively lightweight model that is best adapted to avoid overfitting. Lack of saturation in the test set indicates that PET-MAD could still benefit from an increase in the train set size, and it would be easy, if needed, to obtain a more expressive model by increasing the number of GNN and transformer layers.

\subsection{Ionic transport in lithium thiophosphate}

\begin{figure}
    \centering\includegraphics{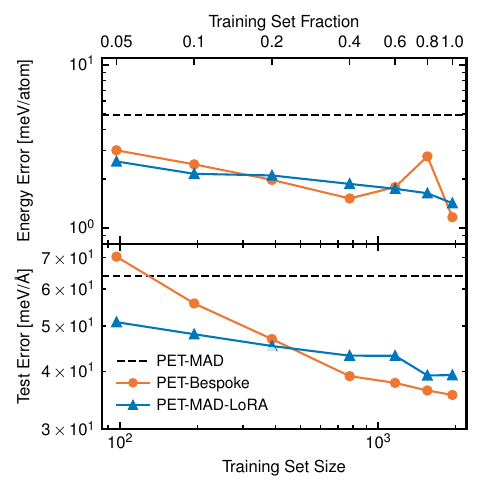}
    \caption{Learning curves (forces) for the Li$_{3}$PS$_{4}$ dataset of Ref.~\cite{Gigli2024}, comparing a bespoke model and a LoRA-finetuned model with rank 8 against the general PET-MAD model. }
\label{fig:LPS-learning-curve}
\end{figure}

Figure \ref{fig:LPS-learning-curve} shows learning curves for the Li$_{3}$PS$_{4}$ dataset for three learning scenarios, i.e., training from scratch (orange), LoRA finetuning with a rank of 8 (blue), and complete fine-tuning of all the model parameters (green). The black dashed line corresponds to the MSE of PET-MAD on the LPS test set. The complete fine-tuned seems to outperform the others of about 20\,\%, possibly because of the reduced number of epochs required during fine-tuning to reach a small error with respect to a model trained from scratch. LoRA finetuning provides slightly higher errors on forces with respect to the bespoke PET model, however from the results in the main text seems that this small difference does not translate in a large difference in the values of the ionic conductivity.

\subsection{Melting point of GaAs}
\begin{figure}
    \centering
    \includegraphics{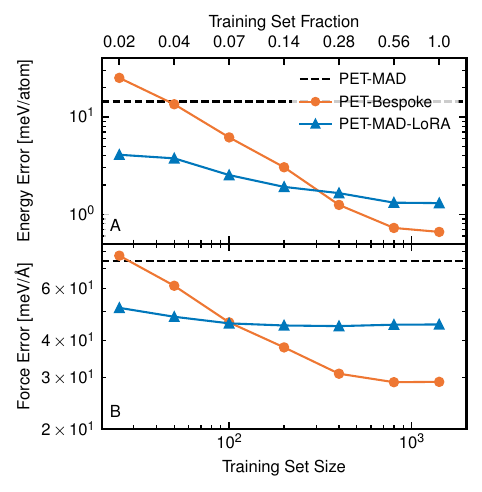}
    \caption{Learning curves (forces) for the Gallium Arsenide dataset, comparing training from scratch (PET-Bespoke) and a LoRA finetuning with LoRA rank 8 against the general PET-MAD model. }
\label{fig:GaAs-learning-curve}
\end{figure}

Figure \ref{fig:GaAs-learning-curve} shows the learning curves for the Gallium Arsenide dataset, for the training of a bespoke PET model and LoRA and fully finetuned PET-MAD models, as well as the base accuracy of PET-MAD (black-dashed line). 

\subsection{Surface segregation in high-entropy alloys}

\begin{figure}
    \centering
    \includegraphics{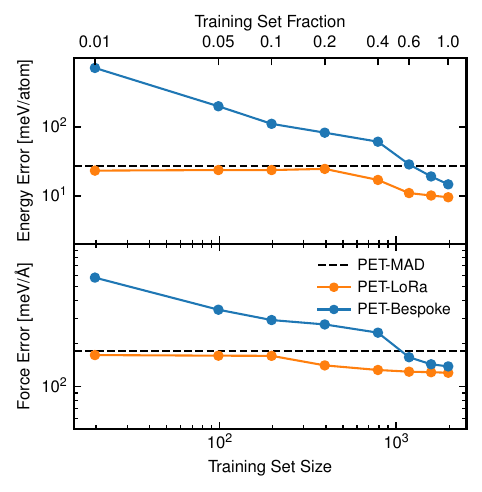}
    \caption{Learning curves (energies) for the subset of the HEA25S dataset \cite{mazi+24jpm}, comparing training from scratch and LoRA finetuning with LoRA rank 8 against the base PET-MAD model predictions error. }
\label{fig:HEA-learning-curve}
\end{figure}

Figure \ref{fig:HEA-learning-curve} shows the learning curves of three models, namely the bespoke PET model trained on a subset of the HEA25S dataset \cite{mazi+24jpm} from scratch, and two fine-tuned models, which start from the pre-trained PET-MAD weights: the fully fine-tuned and a LoRA-finetuned model. Both fine-tuned models show similar training behavior in a low-data regime with a fully funetuned one demonstrating better accuracy as the amount of training data increases. In contrast to other simulations cases done in this work, the bespoke model cannot achieve the same accuracy on a test set, compared to fine-tuned models. This is likely caused by undersampling in the HEA25S subset used for training, which only contains 1975 structures with 25 transition metals, while the original dataset has about 30,000 structures. 

\subsection{Quantum nuclear effects in liquid water}

\begin{figure}
    \centering
    \includegraphics{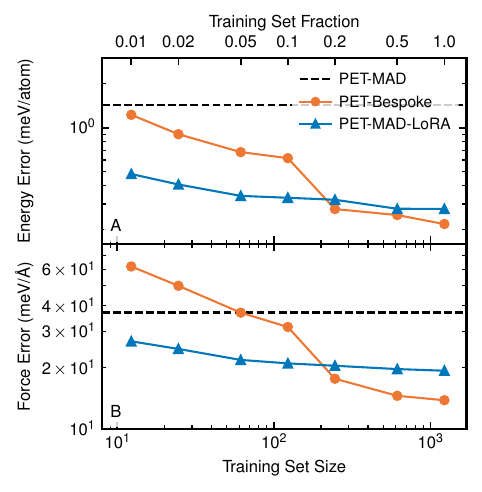}
    \caption{Learning curves for the water dataset, comparing training from scratch, LoRA finetuning with a rank of 8, and LoRA finetuning with a rank of 32 against the base PET-MAD model predictions error. }
    \label{fig:water-ft}
\end{figure}

Figure \ref{fig:water-ft} shows learning curves for the liquid water dataset for three learning scenarios: (1) training from scratch, (2) LoRA finetuning with a rank of 8, (3) LoRA finetuning with a rank of 32. It can be seen that fine-tuning the PET-MAD model under these conditions is advantageous until approximately 20\% of the entire training set (which consists of 1228 structures overall) is used. With more structures, the fine-tuned model is less accurate; this is partially remedied by increasing the rank of LoRA.

\subsection{Quantum nuclear effects in NMR crystallography}
\begin{figure}
    \centering
    \includegraphics{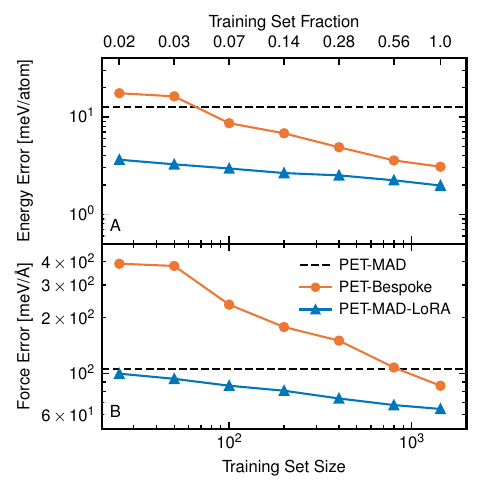}
    \caption{Learning curves (forces) for the Succinic Acid dataset, comparing training from scratch, LoRA finetuning with LoRA rank 8, as well as full finetuning against the general PET-MAD model. }
\label{fig:Succinic-acid-learning-curve}
\end{figure}

Figure \ref{fig:Succinic-acid-learning-curve} shows learning curves for the succinic acid dataset, for all three learning scenarios.

\begin{figure}
    \centering
    \includegraphics{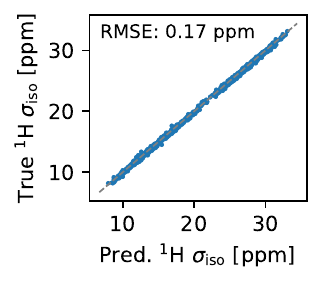}
    \caption{Parity plot of the auxillary model constructed to predict chemical shieldings in succinic acid crystals.}
\label{fig:Succinic-acid-ShiftML-parity}
\end{figure}

For the computation of chemical shielding of succinic acid crystals, we construct an auxilliary model to predict chemical shieldings. We construct one linear model per central species from SOAP descriptors, computed with the featomic library. A parity plot of predicted and reference chemical shielding values is shown in Figure~\ref{fig:Succinic-acid-ShiftML-parity}.

\subsection{Dielectric response of barium titanate}
\begin{figure}
    \centering
    \includegraphics{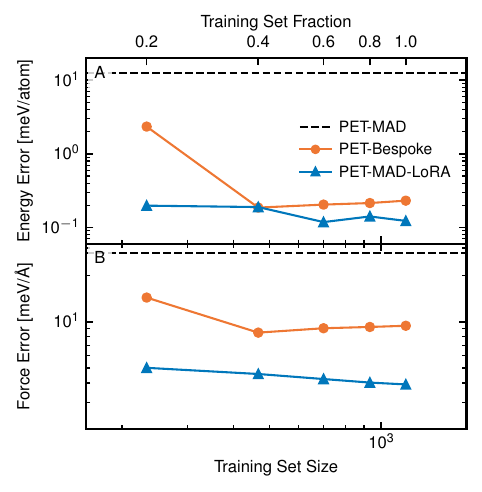}
    \caption{Learning curves for the BTO dataset, comparing training from scratch (PET-Bespoke) and LoRA finetuning with a rank of 8 (PET-MAD-LoRA) against the base PET-MAD model predictions error. }
    \label{fig:batio3-ft}
\end{figure}

Figure \ref{fig:batio3-ft} shows learning curves for the BTO dataset for three learning scenarios, i.e., training from scratch, LoRA finetuning with a rank of 8, and fine-tuning of all the model parameters. Fine-tuned models always outperforms models trained from scratch, possibly because of the reduced number of epochs required during fine-tuning to reach a small error with respect to a model trained from scratch. LoRA finetuning provides slightly smaller errors on forces with respect to full fine-tuning, while at the same time being faster, requiring the update of a reduced number of parameters.

\section{Simulations}
The simulations we perform follow closely the protocol applied in the reference publication. Here we provide an overview of the methodology to explain the technical challenges and provide a few details that can be useful to appreciate the implications of the benchmark accuracy.

\subsection{Ionic transport in lithium thiophosphate}

The ionic conductivities, $\sigma$, in Section \ref{sec:LPS_result} were computed via the Green-Kubo theory of linear response~\cite{Kubo}, which is a practical framework to compute transport coefficients of extended systems~\cite{Gigli2024}. For an isotropic system of $N$ interacting particles reads:
\begin{equation}\label{eq:GKeq}
    \sigma=\frac{\Omega}{3k_{\mathrm{B}}T}\int_0^{\infty} \langle \mathbf{J}_q(\Gamma_t) \cdot \mathbf{J}_q(\Gamma_0) \rangle\, dt,
\end{equation}
where $k_{\mathrm{B}}$ is the Boltzmann constant, $T$ the temperature and $\Gamma_t$ indicates the time evolution of a point in phase space from the initial condition $\Gamma_0$, over which the average $\langle \cdot \rangle$ is performed. $\mathbf{J}_q$ is the charge flux, that depends only on the  velocities of the atoms, $\mathbf{v}_i$, and their charges, $q_i$:
\begin{equation}\label{eq:elec-currents}
    \mathbf{J}_q=\frac{e}{\Omega}\sum_{i} q_i \mathbf{v}_i .
\end{equation}
Here, the sum runs over all the atoms, $e$ is the electron charge, and the $q_i$ are equal to the nominal oxidation number of the atoms. 

\subsection{Melting point of GaAs}

The melting points in Section \ref{sec:GaAs_result} were computed from interface pinning simulations of GaAs liquid solid interface model structures, following Imbalzano's work, studying GaAs with MLIPs \cite{imba-ceri21prm}.
At the melting point, the chemical potential of the liquid phase $\mu_l$ and solid phase $\mu_g$ are identical. Applying a pinning potential restrains a system to coexisting liquid and solid phases separated by a planar interface. A suitable collective variable must be chosen that distinguishes between liquid and solid phase - we will employ the same atom centered Steinhardt Q4 order parameter as chosen by Imbalzano. For a given configuration $A$, the expression for the pinning potential $V_s(A)$, reads:
\begin{equation}
    V_s(A) = k/2 (s(A) - \overline{s})^2
\end{equation}
where, $k$ is the spring constant of the restraint, $s(A)$ the value of the collective variable for a given configuration (the sum over all atom-wise Q4s) and $\overline{s}$ is the value of the collective variable to which the configuration is restrained. When $s(A)$ and $\overline{s}$ are normalized with respect to fully solid and liquid boxes, $\overline{s}$ is set to $\frac{1}{2}$ restraining the system to a half-liquid half-solid box.
The average force of the pinning potential acting on the system is proportional to the chemical potential difference $\Delta\mu $ between the pinned liquid and solid phase at a given temperature:
\begin{equation}
     \Delta\mu \propto k \langle s(A) - \overline{s} \rangle
\end{equation}
The melting point is then determined via root finding by simulating pinned systems at increasing temperatures and determining when the average force is zero.
We compute the collective variable with PLUMED and run molecular dynamics with lammps to determine the average pinning force. Interface structures are constructed by first building elongated   supercells (1152 Atoms, 17 \AA $\times$ 17 \AA $\times$ 90 \AA) of GaAs and partial melting of half of the box, whilst constraining the solid half. These simulations were used to determine the average restraining value $\overline{s}$. The supercells were relaxed with the respective PET potential and then partially melted with NPT MD runs. Production MDs were run for 1~ns, using a 4~fs timestep. The Volume is kept fixed in the x and y directions while a barostat is applied on the z direction to fix the pressure at the interface around 1~bar. Simulations were performed for temperatures from 950~K to 1200~K in intervals of 50~K.

We propagate the model uncertainties of the potential energies to the uncertainties of the melting point via thermodynamic reweighting of the observables computed from the interface pinning simulations that were driven by the respective PET potential. Through the last layer LLPR ensemble uncertainty quantification scheme, we obtain at each timestep of the simulation a committee of $N$ potential energy predictions (where $N=128$ is the number of sampled LLPR ensembles), which are then used to reweight the instantaneous value of the collective variable. The reweighted collective variables are then averaged across the trajectory for each committe member yielding $N$ chemical potential differences. From the chemical potential differences we then compute a set of $N$ melting points, which we then either serve as a nonparametric estimate of the distribution of the computed melting point, or take the standard deviation of the reweighted melting points as an uncertainty estimate of the computed melting point. A more detailed description of using thermodynamic reweighting for uncertainty quantification of thermodynamic averages can be found in Ref.~\cite{imba+21jcp}. 

\subsection{Surface segregation in high-entropy alloys}\label{sec:hea_details}

To study the surface segregation in the CoCrFeMnNi alloy we prepared a surface slab with a \textit{fcc} lattice in the (111) orientation and a $7\times 7\times 11$ supercell containing 539 atoms. Relaxation of both structure and composition of the surface was performed within replica-exchange molecular dynamics run with Monte-Carlo atom swaps with 16 replicas for 200 ps in the NPT ensemble using a 2 fs timestep at zero pressure and logarithmic temperatures grid ranged from 500 K to 1200 K.

We used both the pre-trained PET-MAD and fine-tuned PET-MAD potentials to perform identical REMD/MC runs. The fine-tuning step was performed starting from PET-MAD model weights and training on a subset of the HEA25S dataset \cite{mazi+24jpm} of randomly chosen 2000 structures (1000, 500, 200, 200 and 100 entries from the subsets O, A, B, C and D of the HEA25S dataset, respectively) , recalculated using MAD DFT settings (see Section \ref{sec:dataset_details}). Details of fine-tuning are provided in the Supplementary Information. 

Surface content of the (111) surface of the relaxed CoCrFeMnNi alloy was analyzed if terms of Gibbs surface excess per unit area $\Gamma_a$, which indicates the surface segregation propensity of the elements in the alloy. It is defined as 
\begin{equation}
    \Gamma_a = \frac{N_a - N_a^{B} \cdot N / N^B}{S}, \label{eq:gibbs-excess}
\end{equation}
where $N_a$ and $N_a^B$ correspond to the total number of atoms of element $a$ in the slab, and the number of atoms of element $a$ inside the bulk region of the surface slab. $N$ and $N^B$ represent the total numbers of atoms in the cell and inside the bulk region, respectively, and $S$ is the surface area. The bulk region is defined as a 10\AA{}-thick region around the center of the slab. Therefore, the values of $\Gamma_a > 0$ correspond to enrichment of the surface with element $a$ compared to bulk of the material, while $\Gamma_a < 0$ in contrast correspond to surface depletion. If $\Gamma_a \approx 0$, the concentration of element $a$ at the surface and in the bulk is roughly the same. Defined in this way, $\Gamma_a$ allows the surface composition to be analyzed independently of the choice of bulk layer thickness, providing a macroscopic measure of surface affinity. 

\subsection{Quantum nuclear effects in liquid water}\label{ssec:liquid-water-theory}

The most widespread method to include nuclear quantum effects for equilibrium observables at a constant temperature $T$ is path integral molecular dynamics~\cite{markland2018nuclear} (PIMD). In this variant of molecular dynamics, a number $P$ of equivalent replicas of the system are run simultaneously at temperature $PT$. The system is evolved classically, according to an overall potential which is the sum of the potential energies of the individual replicas, plus a harmonic spring term between all corresponding atoms belonging to adjacent replicas:
\begin{equation}
    V'(\{\mathbf{r}\}_{j=1}^P) = \sum_{j=1}^P V(\mathbf{r}_j) + \frac{1}{2} \omega_P^2 |\Tilde{\mathbf{r}}_j-\Tilde{\mathbf{r}}_{j-1}|^2,
\end{equation}
where $V'(\{\mathbf{r}\}_{j=1}^P)$ is the potential energy of the extended classical system, $j$ indexes the $P$ replicas (also called beads) of the system, and $V(\mathbf{r})$ is the potential energy surface. Furthermore, $\Tilde{\mathbf{r}}$ contains the mass-scaled positions ($\Tilde{r}_i = r_i\sqrt{m_i}$ for every atom $i$), and we define $\Tilde{\mathbf{r}}_0$ so that it ``wraps around'' to $\Tilde{\mathbf{r}}_P$. The angular frequency of the harmonic terms between beads is given by $\omega_P=k_{\mathrm{B}}PT/\hbar$.

The treatment of NQEs for equilibrium observables calculated with PIMD is exact as $P \rightarrow \infty$, and it reduces to that given by classical MD for $P=1$. The calculation of thermodynamic averages from PIMD simulations involves the use of so-called ``estimators''~\cite{markland2018nuclear}, which can range from simple averages over the bead positions for position-dependent observables to more complicated expressions for momentum-dependent observables. In this work, we use the heat capacity estimators from Ref.~\cite{yamamoto2005path}.

All simulations were run with a box of 128 water molecules at constant volume and temperature. The density was chosen to be 997.1 kg/m$^3$ (corresponding to the experimental density of water at 298 K and 1 atm), and the temperature was set to 298 K. 

\subsection{Quantum nuclear effects in NMR crystallography}

We compute thermodynamic averages of chemical shieldings from  MD and PIMD simulations following the exact computational protocol described in Ref.~\cite{enge+21jpcl}, replacing bespoke MLIPs with the PET MAD potentials (Bespoke, PET-MAD-LoRA, PET-MAD) .
Chemical shieldings are computed for the trajectories using bespoke models trained on GIPAW reference calculations for succinic acid crystals.
We fit one linear model considering the local atomic environments of all hydrogen atoms in succinic acid using SOAP descriptors and obtain models of similar accuracy as described in Ref.~\cite{enge+21jpcl}. The $^1$H $\sigma_{\text{iso}}$ test set errors (RMSE) are 0.17 ppm for our linear model compared to 0.16 ppm of the kernel model from the original publication~\cite{enge+21jpcl}. A parity plot of the shielding predictions against the GIPAW reference values can be found in Figure \ref{fig:Succinic-acid-ShiftML-parity}.

\subsection{Dielectric response of barium titanate}

Pressure and temperature along MD simulations are controlled through the Nosé-Hoover barostat and stochastic velocity rescaling, respectively. The system size of 320 atoms corresponds to a $4 \times 4 \times 4$ supercell of the 5-atom BTO unit cell. 
The different structural phases are identified by clustering a reduced-dimensionality representation of the sampled structures with a Gaussian mixture model, which assigns to every sampled structure, $t$, a set of probabilities, $P_k(t)$, corresponding to each identified cluster $k$, effectively labeling the phases~\cite{gigl+22npjcm}. 
The phase transition temperature is estimated by evaluating the relative chemical potential between two phases, $k$ and $k'$:
\begin{align}\label{eq:bto delta mu}
    \Delta \mu^{kk'}(T) = - k_{\mathrm{B}} T \log \frac{\sum_t P_k(t)}{\sum_t P_{k'}(t)}.
\end{align}
At phase coexistence, $ \Delta \mu^{kk'} = 0 $, yielding a practical way to determine the phase transition temperature.
In practice, we perform a linear fit of $\Delta \mu^{kk'}$ as a function of $T$ to locate the temperature where $\Delta \mu^{kk'}(T) = 0$.

The relative static dielectric tensor was computed from the covariance of the cell dipole moment, $\mathbf{M}$, 
\begin{align}\label{eq:dielectric}
    \varepsilon_{r,\alpha\beta} = \delta_{\alpha\beta} + \frac{\operatorname{cov}(M_\alpha, M_\beta)}{\varepsilon_0 \Omega k_{\mathrm{B}} T},
\end{align}
where $ \varepsilon_0 $ is the vacuum permittivity, $ \Omega $ is the system volume, $ k_{\mathrm{B}} $ is the Boltzmann constant, $ \alpha, \beta $ denote Cartesian directions, and $ \delta_{\alpha\beta} $ is the Kronecker delta. In the cubic phase, as identified by the clustering algorithm, the average dipole is assumed to be zero due to centrosymmetry. In this phase, the dielectric tensor is proportional to the identity matrix, effectively reducing to a scalar quantity. As the system transitions from the cubic to the tetragonal phase, the dielectric components parallel and perpendicular to the polarization axis become anisotropic, with dipole fluctuations suppressed along the polarization axis. Upon further cooling and the transition from the tetragonal to orthorhombic phase, all remaining symmetries are broken (except for those under permutation of Cartesian directions), resulting in six independent components.

The cell dipole moments are obtained by evaluating an equivariant $\lambda$-SOAP-based linear model trained on the dipole dataset from Ref.~\cite{gigl+22npjcm}. These dipoles are then used to compute the dielectric tensors using Eq.~\eqref{eq:dielectric}. The parallel and perpendicular components, used for plotting the dielectric tensor in the tetragonal phase, are computed by projecting the dielectric tensor onto the dipole vector and its orthogonal complement, respectively:
\begin{align}
    \varepsilon_\parallel &= \sum_{\alpha\beta} \varepsilon_{r,\alpha\beta} \frac{M_{\alpha}M_{\beta}}{|\mathbf{M}|^2}, \\
    \varepsilon_\perp &= \sum_{\alpha\beta} \varepsilon_{r,\alpha\beta} \left(\delta_{\alpha\beta} - \frac{M_{\alpha}M_{\beta}}{|\mathbf{M}|^2}\right).
\end{align}

\section{Non-conservative MD}\label{sec:directforces}

We perform molecular-dynamics simulations of the molten phase of the ionic liquid 
1-Butyl-3-methylimidazolium chloride (BMIM-Cl), starting from configurations equilibrated at $500$~K. 
We compare simulations using (i) the standard, conservative forces computed as derivatives of the PET-MAD energy head; (ii) the non-conservative, direct force prediction head (which is approximately 2 times faster); (iii) a multiple-time-stepping
setup in which the conservative forces are evaluated every 8 steps to correct the non-conservative trajectory (which is approximately 1.8 times faster than conservative MD).
We refer the reader to Ref.~\citenum{Bigi2024} for a general discussion of the advantages and problems of direct force prediction. 

\begin{figure*}[tbh]
    \centering
    \includegraphics[width=1.0\linewidth]{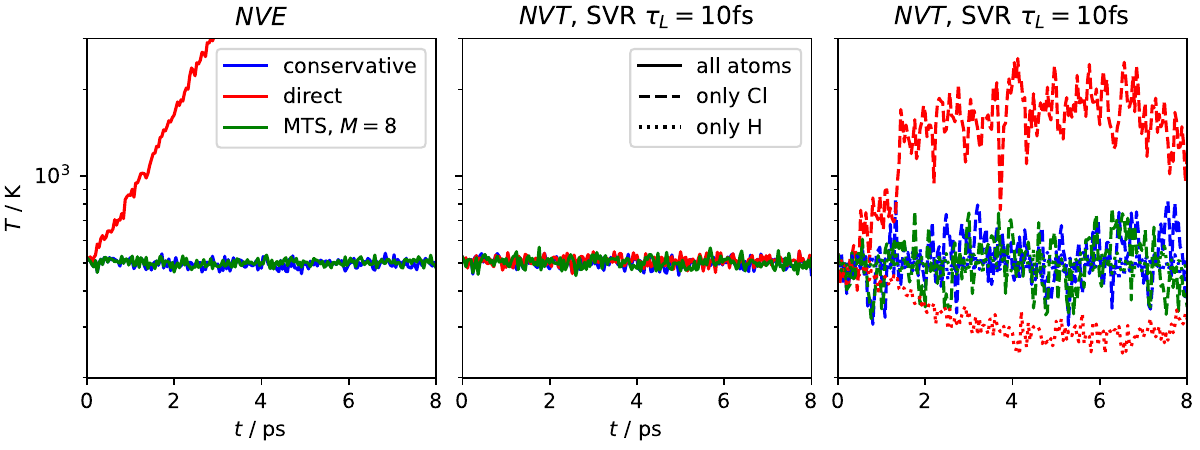}
    \caption{Kinetic temperature profiles along MD simulations, using conservative force models (blue), direct force heads (red) and multiple time stepping (green). (left) Constant-energy simulations initialized from a structure equilibrated at 500~K; note the exponential drift occurring with non-conservative forces. (middle) Constant-temperature simulations using stochastic velocity rescaling, with a thermostat relaxation time $\tau_L=$10~fs; direct-force simulations reach a steady-state with the overall temperature fluctuating around the target $T=500$~K. (right) Temperature profiles for individual atomic types (Cl and H shown) demonstrate that the direct-force simulations are problematic even with aggressive global thermostatting; Cl reach a steady-state temperature above 2000~K, which is compensated by H (and other) atoms being well below the average temperature. Note that the larger kinetic energy fluctuations for Cl atoms, seen also for conservative trajectories, is normal and consistent with the fact that only a few Cl ions are present in the simulation box. }
    \label{fig:directforce-temperature}
\end{figure*}

\begin{figure*}[tbh]
    \centering
    \includegraphics[width=0.7\linewidth]{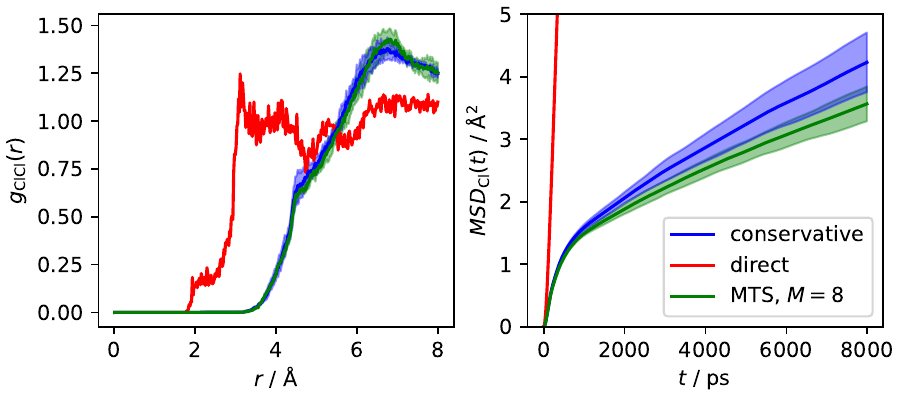}
    \caption{Cl-Cl pair correlation function (left) and Cl mean square displacement (right) for simulations of BMIM-Cl at $T=500$~K, using conservative force models (blue), direct force heads (red) and multiple time stepping (green). 
    }
    \label{fig:directforce-properties}
\end{figure*}

We first perform constant-energy simulations, using a time step of 0.5~fs and a velocity Verlet integrator. 
When using a conservative force calculation, the kinetic temperature of the atoms as a function of time (Fig.~\ref{fig:directforce-temperature}a) fluctuates around the initial equilibrium temperature, as expected. 
The non-conservative head leads to fast, catastrophic drift of the kinetic energy, that quickly leads to molecular dissociation and completely unphysical outcomes. 
A multiple-time-stepping strategy, instead, yields a stable trajectory, that is entirely consistent with  conservative MD.
As shown for other non-conservative potentials in Ref.~\citenum{Bigi2024}, using an aggressive global thermostat avoids a drift in the overall temperature, but breaks energy conservation, leading to different chemical species (or more broadly, different degrees of freedom) reaching a different steady-state kinetic energy
(Fig.~\ref{fig:directforce-temperature}b). 
Even in this case, a multiple-time-stepping simulation avoids these artifacts, while retaining most of the computational advantage.

Multiple time step calculations also yield static and dynamic properties that are consistent with those of a conservative calculation, as shown in Fig.~\ref{fig:directforce-properties}. 
Both the Cl-Cl correlation function (reporting on the local structure in the liquid) and the Cl diffusion (reporting on the ionic conductivity) are within the statistical uncertainty. On the other hand, direct-force trajectories lead to extremely fast Cl diffusion (consistent with the much higher steady-state temperature) and the pair correlation function show the association of Cl dimers, another clear failure of the non-conservative sampling.

\end{document}